\documentclass[journal, letterpaper]{IEEEtran}
\usepackage[T1]{fontenc}

\newcommand{\maximize}{\mathop{\rm maximize}}
\newcommand{\minimize}{\mathop{\rm minimize}}

\newtheorem{theorem}{Theorem}
\newtheorem{remark}{Remark}
\newtheorem{lemma}{Lemma}

\newtheorem{assumption}{Assumption}
\newtheorem{definition}{Definition}

\newcommand{\argmax}{\mathop{\rm argmax}}

\usepackage{cite}
\ifCLASSINFOpdf
\usepackage[pdftex]{graphicx}
\usepackage[unicode,hidelinks]{hyperref}
\else
\usepackage[dvips]{graphicx}
\DeclareGraphicsExtensions{.eps}
\fi
\usepackage[cmex10]{amsmath}
\usepackage{amssymb}
\usepackage{mathtools}
\usepackage[ruled]{algorithm}
\usepackage{algpseudocode}
\usepackage{mathrsfs}
\usepackage[caption=false,font=footnotesize]{subfig}
\usepackage{fixltx2e}
\usepackage{makecell}
\usepackage{multirow}
\usepackage{hhline}
\usepackage{enumitem}
\usepackage{microtype}
\usepackage{color}
\usepackage[cmintegrals]{newtxmath}
\ifCLASSOPTIONcaptionsoff
\usepackage[nomarkers]{endfloat}
\let\MYoriglatexcaption\caption
\renewcommand{\caption}[2][\relax]{\MYoriglatexcaption[#2]{#2}}
\fi
\usepackage{kotex}
\usepackage{setspace}
\usepackage{etoolbox}
\AtBeginEnvironment{algorithmic}{\onehalfspacing}

\newcommand{\bh}{\ensuremath{{\mathbf h}}}
\newcommand{\bg}{\ensuremath{{\mathbf g}}}

\newcommand{\ba}{\ensuremath{{\mathbf a}}}

\newcommand{\bv}{\ensuremath{{\mathbf v}}}

\newcommand{\bs}{\ensuremath{{\mathbf s}}}
\newcommand{\bm}{\ensuremath{{\mathbf m}}}
\newcommand{\bx}{\ensuremath{{\mathbf x}}}

\newcommand{\bn}{\ensuremath{{\mathbf n}}}

\newcommand{\bw}{\ensuremath{{\mathbf w}}}

\newcommand{\btheta}{\ensuremath{{\boldsymbol \uptheta}}}

\newcommand{\calU}{\ensuremath{\mathcal{U}}}

\newcommand{\calS}{\ensuremath{\mathcal{S}}}
\newcommand{\calA}{\ensuremath{\mathcal{A}}}

\makeatletter
\newcommand{\algmargin}{\the\ALG@thistlm}
\makeatother
\newlength{\whilewidth}
\algdef{SE}[parWHILE]{parWhile}{EndparWhile}[1]
{\parbox[t]{\dimexpr\linewidth-\algmargin}{%
		\hangindent\whilewidth\strut\algorithmicwhile\ #1\ \algorithmicdo\strut}}{\algorithmicend\ \algorithmicwhile}%
\algnewcommand{\parState}[1]{\State%
	\parbox[t]{\dimexpr\linewidth-\algmargin}{\setstretch{0.9}\strut #1 \strut\vspace{0.05in}}}

\begin{document}
\title{Adaptive Transmission Scheduling in Wireless Networks for Asynchronous Federated Learning}

\author{Hyun-Suk~Lee and Jang-Won Lee, \textit{Senior Member, IEEE}
	\thanks{This work was supported in part by and the National Research Foundation of Korea (NRF) grant through the Korea Government (MSIT) under Grant 2021R1G1A1004796, and in part by the NRF grant through the Korea Government (MSIT) under Grant 2019R1A2C2084870. \textit{(Corresponding author: Jang-Won Lee)}
	
	H.-S. Lee is with the School of Intelligent Mechatronics Engineering, Sejong University, Seoul 05006, South Korea (e-mail: hyunsuk@sejong.ac.kr) and J.-W. Lee is with the Department of Electrical and Electronic Engineering, Yonsei University, Seoul 03722, South Korea (e-mail: jangwon@yonsei.ac.kr).}
}

\maketitle

\begin{abstract}
In this paper, we study asynchronous federated learning (FL) in a wireless distributed learning network (WDLN). To allow each edge device to use its local data more efficiently via asynchronous FL, transmission scheduling in the WDLN for asynchronous FL should be carefully determined considering system uncertainties, such as time-varying channel and stochastic data arrivals, and the scarce radio resources in the WDLN.
To address this, we propose a metric, called an effectivity score, which represents the amount of learning from asynchronous FL. We then formulate an Asynchronous Learning-aware transmission Scheduling (ALS) problem to maximize the effectivity score and develop three ALS algorithms, called ALSA-PI, BALSA, and BALSA-PO, to solve it. 
If the statistical information about the uncertainties is known, the problem can be optimally and efficiently solved by ALSA-PI. Even if not, it can be still optimally solved by BALSA that learns the uncertainties based on a Bayesian approach using the state information reported from devices. 
BALSA-PO suboptimally solves the problem, but it addresses a more restrictive WDLN in practice, where the AP can observe a limited state information compared with the information used in BALSA. We show via simulations that the models trained by our ALS algorithms achieve performances close to that by an ideal benchmark and outperform those by other state-of-the-art baseline scheduling algorithms in terms of model accuracy, training loss, learning speed, and robustness of learning. These results demonstrate that the adaptive scheduling strategy in our ALS algorithms is effective to asynchronous FL.
\end{abstract}


\section{Introduction}
\label{sec:introduction}
Nowadays, a massive amount of data is generated from devices, such as mobile phones and wearable devices, which can be used for a wide range of machine learning (ML) applications from healthcare to autonomous driving. As the computational and storage capabilities of such distributed devices keep growing, distributed learning has become attractive to efficiently exploit the data from devices and to address privacy concerns. Due to the emergence of the need for distributed learning, federated learning (FL) has been widely studied as a potentially viable solution for distributed learning \cite{mcmahan2017communication,lim2020federated}.
FL allows to learn a shared central model from the locally trained models of distributed devices under the coordination of a central server while the local training data at the devices is not shared with the central server.

In FL, the central server and the devices should communicate with each other to transmit the trained models.
For this, a wireless network composed of one AP and multiple devices has been widely considered \cite{amiri2020federated,amiri2020machine,yang2019scheduling,wang2019adaptive,amiriconvergence,shi2020joint,chen2020joint,xu2020client,xia2020multi}, and the AP in the network plays the role of the central server in FL.
FL is operated in multiple rounds. In each round, first, the AP broadcasts the current central model. In typical FL, \textit{all} devices substitute their local models into the received central one and train them using their local training data.
Then, the locally trained model of each device is uploaded to the AP and the AP aggregates them into the central model.
This enables training the central model in a distributed manner while avoiding privacy concerns because it does not require each device to upload its local training data.

However, as FL has been implemented in wireless networks, the limited radio resources of wireless networks have been raised as a critical issue since the radio resources restrict the number of devices that can upload the local models to the AP simultaneously \cite{shi2020joint,amiriconvergence}.
To address the issue, in \cite{xu2020client,yang2019scheduling,wang2019adaptive,amiriconvergence,shi2020joint,chen2020joint,xia2020multi}, device scheduling procedures for FL have been proposed in which the AP schedules the devices to train their local models and upload them considering the limited radio resources.
Then, in each round, only the scheduled devices do so and the AP aggregates the local models uploaded from the scheduled devices.
On the other hand, the not-scheduled devices do not train their local models and their local data that have been arrived during the round is stored for future use.
Hence, every locally trained model in each round is synchronously aggregated into the central model in the round (i.e., once a device trains its local model in a round, it must be aggregated into the central model in that round and its use in future rounds is not allowed), and such FL is called \textit{synchronous} FL in the literature.

Device scheduling procedures in synchronous FL have effectively addressed the issue of the limited radio resources in wireless networks, but at the same time, they may cause the inefficiency of utilizing local computation resources and the loss of local data due to too much pileup \cite{chen2019communication,chen2019asynchronous}.
These issues are raised because of the not-scheduled devices in each round that do not train their local models.
Hence, to address them, we can allow even the not-scheduled devices to train their local models and store them for future use.
This enables each device to continually train its local model by using the arriving local data in an online manner, which avoids waste of local computation resources and too much pileup of the local data \cite{chen2019communication,chen2019asynchronous}.
However, at the same time, when aggregating the stored locally trained models, it causes the time lag between the stored local models and the current central model due to the asynchronous local model aggregation.
In the literature, such FL with the time lag is called \textit{asynchronous} FL.
In addition, such devices who store the previous locally trained models are typically called \textit{stragglers}, and when aggregating them into the central model, they may cause an adverse effect to the convergence of the central model because of the time lag \cite{chen2019asynchronous}.

Recently, in the ML literature, several works on asynchronous FL have addressed the harmful effects from the stragglers which inevitably exist due to network circumstances such as time-varying channels and scarce radio resources \cite{chen2019asynchronous,zheng17b,chen2019communication,xie2019asynchronous}.
They introduced various approaches which address the stragglers when updating the central model and the local models; balancing between the previous local model and current one \cite{xie2019asynchronous,chen2019asynchronous}, adopting dynamic learning rates \cite{zheng17b,chen2019asynchronous,chen2019communication}, and using a regularized loss function \cite{xie2019asynchronous,chen2019asynchronous}.
However, they mainly focused on addressing the incurred stragglers and did not take into account the issues relevant to the implementation of FL, which are related to the occurrence of the stragglers.
Hence, it is necessary to study an asynchronous FL procedure in which the key challenges of implementing asynchronous FL in a wireless network are carefully considered.

When implementing FL in the wireless network, the scarcity of radio resources is one of the key challenges. Hence, several existing works focus on reducing the communication costs of a single local model transmission of each edge device; analog model aggregation methods using the inherent characteristic of a wireless medium \cite{amiri2020machine,amiri2020federated}, a gradient reusing method \cite{chen2018lag}, and local gradient compressing methods \cite{lin2018deep,xu2020ternary}.
Meanwhile, other existing works in \cite{wang2019adaptive,yang2019scheduling,shi2020joint,amiriconvergence,xu2020client,chen2020joint,xia2020multi}
focus on scheduling the transmission of the local models for FL in which only scheduled edge devices participate in FL.
Since typical scheduling strategies in wireless networks \cite{li2020multi,lee2019resource} do not consider FL, for effective learning, various scheduling strategies for FL have been proposed based on traditional scheduling policies \cite{yang2019scheduling}, scheduling criteria to minimize the training loss \cite{wang2019adaptive,chen2020joint,amiriconvergence,shi2020joint}, an effective temporal scheduling pattern to FL \cite{xu2020client}, and multi-armed bandits \cite{xia2020multi}. They allow the AP to efficiently use the radio resources to accelerate FL.

However, these existing works on FL implementation in the wireless network have been studied only for synchronous FL, and do not consider the characteristics of asynchronous FL, such as the stragglers and the continual online training of the local models, at all.
Nevertheless, the existing methods on reducing the communication costs can be used for asynchronous FL since in asynchronous FL, the edge devices should transmit their local model to the AP as in synchronous FL.
On the other hand, the methods on scheduling the transmission of the local models may cause a significant inefficiency of learning if they are adopted to asynchronous FL.
This is because most of them consider each individual FL round separately and do not address transmission scheduling over multiple rounds.
In asynchronous FL, each round is highly inter-dependent since the current scheduling affects the subsequent rounds due to the use of the stored local models from the stragglers.
Hence, the transmission scheduling for asynchronous FL should be carefully determined considering the stragglers over multiple rounds.
In addition, in such scheduling over multiple rounds, the effectiveness of learning depends on system uncertainties, such as time-varying channels and stochastic data arrivals.
In particular, the stochastic data arrivals become more important in asynchronous FL since the data arrivals are directly related to the amount of the straggled local data due to the continual online training.
However, the existing works do not consider stochastic data arrivals at edge devices.

\setlength\tabcolsep{4 pt}
\renewcommand{\arraystretch}{1.0}
\begin{table}[!t]
\fontsize{8pt}{1em}\selectfont
	\centering
	\caption{Comparison of Our Work and Related Works on Transmission Scheduling for FL ($\surd$: Considered / $\times$: Not Considered)}
	\label{table:related_works}
	\begin{tabular}{c|c|c|c|c}
		\hline
		 & \makecell{\textbf{Wireless}\\\textbf{channel}} & \makecell{\textbf{Multiple}\\\textbf{rounds}} & \makecell{\textbf{Stochastic}\\\textbf{data arrivals}} & \makecell{\textbf{Stragglers from}\\\textbf{async. FL}} \\\hline\hline
		 \cite{wang2019adaptive,chen2020joint,amiriconvergence,shi2020joint,yang2019scheduling} & $\surd$ & $\times$ & $\times$ & $\times$ \\\hline
\cite{xu2020client,xia2020multi}  & $\surd$ & $\surd$ & $\times$ & $\times$ \\\hline
		 Our work & $\surd$ & $\surd$ & $\surd$ & $\surd$ \\\hline
	\end{tabular}
\end{table}

In this paper, we study asynchronous FL considering the key challenges of implementing it in a wireless network.
To the best of our knowledge, our work is the first to study asynchronous FL in a wireless network.
Specifically, we propose an asynchronous FL procedure in which the characteristics of asynchronous FL, time-varying channels, and stochastic data arrivals of the edge devices are considered for transmission scheduling over multiple rounds.
The comparison of our work and the existing works on transmission scheduling in FL is summarized in Table \ref{table:related_works}.
We then analyze the convergence of the asynchronous FL procedure.
To address scheduling the transmission of the local models in the asynchronous FL procedure, we also propose a metric called an \textit{effectivity score}. It represents the amount of learning from asynchronous FL considering the properties of asynchronous FL including the harmful effects on learning due to the stragglers.
We formulate an asynchronous learning-aware transmission scheduling (ALS) problem to maximize the effectivity score while considering the system uncertainties (i.e., the time-varying channels and stochastic data arrivals).
We then develop the following three ALS algorithms that solve the ALS problem:
\begin{itemize}
\item First, an ALS algorithm with the perfect statistical information about the system uncertainties (ALSA-PI) optimally and efficiently solves the problem using the state information reported from the edge devices in the asynchronous FL procedure and the statistical information.
\item Second, a Bayesian ALS algorithm (BALSA) solves the problem using the state information without requiring any a priori information. Instead, it learns the system uncertainties based on a Bayesian approach. We prove that BALSA is optimal in terms of the long-term average effectivity score by its regret bound analysis.
\item Third, a Bayesian ALS algorithm for a partially observable WDLN (BALSA-PO) solves the problem only using partial state information (i.e., channel conditions). It addresses a more restrictive WDLN in practice, where each edge device is allowed to report only its current channel condition to the AP.
\end{itemize}
Through experimental results, we show that ALSA-PI and BALSAs (i.e., BALSA and BALSA-PO) achieve performance close to an ideal benchmark with no radio resource constraints and transmission failure.
We also show that they outperform other baseline scheduling algorithms in terms of training loss, test accuracy, learning speed, and robustness of learning.\looseness=-1

The rest of this paper is organized as follows. Section II introduces a WDLN with asynchronous FL. In Section III, we formulate the ALS problem considering asynchronous FL. In Section IV, we develop ALSA-PI, BALSA, and BALSA-PO, and in Section V, we provide
experimental results. Finally, we conclude in Section VI.

\section{Wireless Distributed Learning Network with Asynchronous FL}
In this section, we introduce typical learning strategies of asynchronous FL provided in \cite{xie2019asynchronous,zheng17b,chen2019asynchronous,chen2019communication}.
We then propose an asynchronous FL procedure to adopt the learning strategies in a WDLN.
For ease of reference, we summarize some notations in Table \ref{table:notations}.

\setlength\tabcolsep{4 pt}
\renewcommand{\arraystretch}{1.1}
\begin{table}[!t]
\fontsize{8pt}{1em}\selectfont
	\centering
	\caption{List of notations}
	\label{table:notations}
	\begin{tabular}{cp{7cm}}
		\hline
		\textbf{Notation} & \textbf{Description} \\
		\hline
		$\bw^t$ & Central parameters of the AP in round $t$ \\
		$\bw_u^t$ & Local parameters of device $u$ in round $t$ \\		
		$c_u^t$ & Central learning weight of device $u$ in round $t$ \\
		$n_u^t$ & Number of aggregated samples of device $u$ from the latest successful local update transmission to at the beginning of round $t$ \\
		$m_u^t$ & Number of arrived samples of device $u$ during round $t$ \\
		$N_u^t$ & Total number of samples from device $u$ used for the central updates before the beginning of round $t$ \\
		$\Delta_u^t$ & Local update of device $u$ in round $t$ \\
		$\eta_u^t$ & Local learning rate of device $u$ in round $t$ \\
		$\psi_u^t$ & Local gradient of device $u$ in round $t$ \\
		$h_u^t$ & Channel gain of device $u$ in round $t$ \\
		$a_u^t$ & Transmission scheduling indicator of device $u$ in round $t$ \\
		$x_u^t$ & Successful transmission indicator of device $u$ in round $t$ \\
		$\btheta_u^C$ & System parameters related to the channel gain of device $u$ \\
		$\theta_u^P$ & System parameter related to the sample arrival of device $u$ \\
		$\mu^1$ & Prior distribution for the parameters $\btheta$ \\
		$\mu^t$ & Posterior distribution for the parameters $\btheta$ in round $t$ \\
		$t_k$ & Start time of stage $k$ of BALSA \\
		$T_k$ & Length of stage $k$ of BALSA \\
		\hline
	\end{tabular}
\end{table}

\vspace{-0.1in}
\subsection{Central and Local Parameter Updates in Asynchronous FL}
\label{sec:parameter_update}
Here, we introduce typical learning strategies to update the central and local parameters in asynchronous FL \cite{xie2019asynchronous,zheng17b,chen2019asynchronous,chen2019communication}, which address the challenges due to the stragglers.
To this end, we consider one access point (AP) that plays a role of a central server in FL and $U$ edge devices.
The set of edge devices is defined as $\mathcal{U}=\{1,2,...,U\}$.
In asynchronous FL, an artificial neural network (ANN) model composed of multiple parameters is trained in a distributed manner to minimize an empirical loss function $l(\bw)$, where $\bw$ denotes the parameters of the ANN.
Asynchronous FL proceeds over a discrete-time horizon composed of multiple rounds.
The set of rounds is defined as $\mathcal{T}=\{1,2,...\}$ and the index of rounds is denoted by $t$.
Then, we can formally define the problem of asynchronous FL with a given local loss function at device $u$, $f_u$, as follows:\looseness=-1
\begin{equation}
\label{eqn:problem}
\minimize_{\bw} ~l(\bw) = \frac{1}{K}\sum_{u\in\mathcal{U}} \sum_{k=1}^{K_u} f_{u}(\bw,k),
\end{equation}
where $K_u$ denotes the number of data samples of device $u$,\footnote{For simple presentation, we omit the word ``edge'' from edge device in the rest of the paper.} $K=\sum_{u=1}^U K_u$, and $f_{u}(\bw,k)$ is an empirical local loss function defined by the parameters $\bw$ at $k$-th data sample of device $u$.
To solve this problem, in asynchronous FL, every device trains parameters by using its locally available data in an online manner for efficient learning on each device and avoiding too much pileup of the local data.
Then, partial devices are scheduled to transmit the trained parameters to the AP.
By using the received parameters from the scheduled devices, the AP updates its parameters.
We call the parameters at the AP central parameters and those at each device local parameters.
The details of how to update the local and central parameters in asynchronous FL are provided in the following.

\subsubsection{Local Parameter Updates}
\label{sec:local_update}
We describe local parameter updates at each device in asynchronous FL.
Let the central parameters of the AP in round $t$ be denoted by $\bw^t$.
Note that the central parameters $\bw^t$ is calculated by the AP in round $t-1$, which will be described in the central parameter updates section later.
In round $t$, the AP first broadcasts the central parameters $\bw^t$ to all devices and each device $u$ replaces its local parameters $\bw_u^t$ with $\bw^t$.
Thus, $\bw^t$ becomes the initial parameters of local training at each device in round $t$.
Then, the device trains its local parameters $\bw_u^t$ using its local data samples that have arrived since its previous local parameter update.
To this end, it applies a gradient-based update method using the regularized loss function defined as follows \cite{xie2019asynchronous,chen2019asynchronous}:\looseness=-1
\begin{equation}
s_u(\bw_u^t)=f_u(\bw_u^t)+\frac{\lambda}{2}||\bw_u^t-\bw^t||,
\end{equation}
where the second term  mitigates the deviations of the local parameters from the central ones and $\lambda$ is the parameter for the regularization.
We denote the local gradient of device $u$ calculated using the local parameters $\bw_u^t$ and its local data samples in round $t$ by $\nabla g_u^t$.
In asynchronous FL, the local gradient which has not been transmitted in the previous rounds will be aggregated to the current local gradient.
Such local gradients not transmitted in the previous rounds are called delayed local gradients.
In the literature, the devices who have such delayed local gradients are called \textit{stragglers}.
It has been shown that they adversely affect model convergence since the parameters used to calculate the delayed local gradients are different from the current local parameters used to calculate the current local gradients \cite{smith2017federated,xie2019asynchronous,chen2019asynchronous}.

To address this issue, when aggregating the previous local gradients and current ones, we need to balance between them. To this end, in \cite{xie2019asynchronous,chen2019asynchronous}, the decay coefficient $\beta$ is introduced and used when aggregating the local gradients as
\begin{equation}
\label{eqn:local_gradient_aggregating}
\psi_u^t=\nabla g_u^t + \beta\psi_u^{t-1},
\end{equation}
where $\psi_u^t$ is the aggregated local gradient of device $u$ in round $t$.
By using the aggregated local gradient, we define the local \textit{update} of device $u$ in round $t$ as
\begin{equation}
\label{eqn:local_update}
\Delta_u^t=\eta_u^t\psi_u^t,
\end{equation}
where $\eta_u^t$ is the local learning rate of the local parameters of device $u$ in round $t$.
It is worth emphasizing that each device uploads its local \textit{update} to the AP for updating the central parameters if scheduled.
Moreover, a dynamic local learning rate has been widely applied to address the stragglers.
The dynamic learning rate of device $u$ in round $t$ is determined as follows  \cite{baytas2016asynchronous,chen2019communication,chen2019asynchronous}:
\begin{equation}
\label{eqn:dynamic_learning_rate}
\eta_u^t=\eta_d\max\left\lbrace 1,\log\left(d_u^t\right) \right\rbrace,
\end{equation}
where $\eta_d$ is an initial value of the local learning rate and $d_u^t$ is the number of delayed rounds since the latest successful local update transmission of device $u$ in round $t$.
Each device transmits its local update $\Delta_u^t$ to the AP according to the transmission scheduling, and then, updates its aggregated local gradient according to the local update transmission as
\begin{equation}
\label{eqn:gradient_balancing}
\psi_u^t=\begin{cases} 0, &\!\!\textrm{if device $u$ successfully transmit its local update} \\
\psi_u^{t}, &\!\!\textrm{otherwise.} \end{cases}
\end{equation}

\subsubsection{Central Parameter Updates}
\label{sec:central_update}
In round $t$, device $u$ obtains its local update, $\Delta_u^t$, by using its local data and delayed local gradients as in \eqref{eqn:local_gradient_aggregating} and \eqref{eqn:local_update}, respectively.
After the scheduled devices transmit their local updates to the AP, the AP recalculates its central parameters by aggregating the successfully received local updates from the scheduled devices.
To represent this, we define the set of the devices whose local updates are successfully received at the AP in round $t$ by $\bar{\calU}^t$.
Then, the central parameters are updated as follows \cite{chen2019asynchronous,chen2019communication}:
\begin{equation}
\bw^{t+1}=\bw^t-\sum_{u\in\bar{\calU}^t} c_u^t\Delta_u^t, \label{eqn:parameter_update}
\end{equation}
where $c_u^t$ is the central learning weight of device $u$ in round $t$.
It is worth empasizing that $\bw^{t+1}$ is calculated by the AP in round $t$ and will be broadcasted to all devices in round $t+1$.
The central learning weight of each device is determined according to the contribution of the device to the central parameter updates so far, which is evaluated according to the number of the samples used for the central parameter updates from the device \cite{amiri2020federated,amiri2020machine,yang2019scheduling,wang2019adaptive,amiriconvergence,shi2020joint,chen2020joint,xia2020multi,xu2020client,chen2019asynchronous,zheng17b,chen2019communication,xie2019asynchronous}.
We define the total number of samples from device $u$ used for the central updates before the beginning of round $t$ as $N_u^t$,
the number of aggregated samples of device $u$ from the latest successful local update transmission to the beginning of round $t$ as $n_u^t$,
and the number of samples of device $u$ that have arrived during round $t$ as $m_u^t$.
Then, the central learning weight of device $u$ in round $t$ is determined as
\begin{equation}
c_u^t=\frac{N_u^t+n_u^t+m_u^t}{\sum_{u\in\calU}N_{u}^t+\sum_{u\in\bar{\calU}^t}n_u^t+m_u^t}.
\end{equation}

\subsection{Asynchronous FL Procedure in WDLN}
\label{sec:WDLN}
We consider a WDLN consisting of one AP and $U$ devices that cooperatively operate FL over the discrete-time horizon composed of multiple rounds that have an identical time duration.
In the WDLN, the local data stochastically and continually arrives at each device and the device trains its local parameters in each round by using the arrived data.
Due to the scarce radio resources of the WDLN, it is impractical that all devices transmit their local parameters to the AP simultaneously.
Hence, FL in the WDLN becomes asynchronous FL and the devices who will transmit their local parameters in each round should be carefully scheduled by the AP for training the central parameters effectively.\looseness=-1

We now propose an asynchronous FL procedure for the WDLN in which the central and local parameter update strategies introduced in the previous subsection are adopted.
In the procedure, the characteristics of the WDLN, such as time-varying channel conditions and system bandwidth, are considered as well.
This allows us to implement asynchronous FL in the WDLN while addressing the challenges due to the scarcity of radio resources in the WDLN.
We denote the channel gain of device $u$ in round $t$ by $h^t_u$.
We assume that the channel gain of device $u$ in each round follows a parameterized distribution and denote the corresponding system parameters by $\btheta_u^C$.
For example, for Rician fading channels, $\btheta_u^C$ represents the parameters of the Rician distribution, $\Omega$ and $K$.
The vector of the channel gain in round $t$ is defined by $\bh^t=\{h_u^t\}_{u\in\calU}$.

In the procedure, each round $t$ is composed of three phases: transmission scheduling phase, local parameter update phase, and parameter aggregation phase.
In the transmission scheduling phase, the AP observes the state information of the devices which can be used to determine the transmission scheduling.
In specific, we define the state information in round $t$ as $\bs^t=(\bh^t,\bn^t)$, where $\bn^t=\{n_u^t\}_{u\in\calU}$ is the vector of the number of aggregated samples of each device from its latest successful local update transmission to the beginning of round $t$.
It is worth noting that in the following sections, we consider a \textit{partially observable} WDLN as well, in which the AP can observe the channel gain $\bh^t$ only, to address more restrictive environments in practice.
Then, the AP determines the transmission scheduling for asynchronous FL based on the state information.
In the local parameter update phase, the AP broadcasts its central parameters $\bw^t$ to all devices.
We do not consider the transmission failure for $\bw^t$ at each device as in the related works \cite{chen2020joint,amiriconvergence,shi2020joint,xu2020client}, i.e., all devices receive $\bw^t$ successfully.
This is reasonable because the probability of the transmission failure for $\bw^t$ is small in practice since the AP can reliably broadcast the central parameters by using much larger transmission power than that of the devices and robust transmission methods considering the worst-case device.%
\footnote{It is worth noting that this assumption is only to avoid unnecessarily complicated modeling. Besides, our proposed method can be easily used even without this assumption as well by allowing each device to maintain its previous local parameters when its reception of the central parameters is failed.}
Each device $u$ replaces its local parameters with the received central parameters (i.e., $\bw_u^t=\bw^t$). Then, it trains the parameters using its local data samples which have arrived since the local parameter update phase in the previous round and calculates the local update $\Delta_u^t$ in \eqref{eqn:local_update} as described in Section \ref{sec:local_update}. Finally, in the parameter aggregation phase, each scheduled device transmits its local update to the AP. Then, the AP updates the central parameters by averaging them as in \eqref{eqn:parameter_update}.
In the following, we describe the asynchronous FL procedure in the WDLN in more detail.\looseness=-1

\begin{figure*}[!t]
	\centering
	\includegraphics[width=0.8\linewidth]{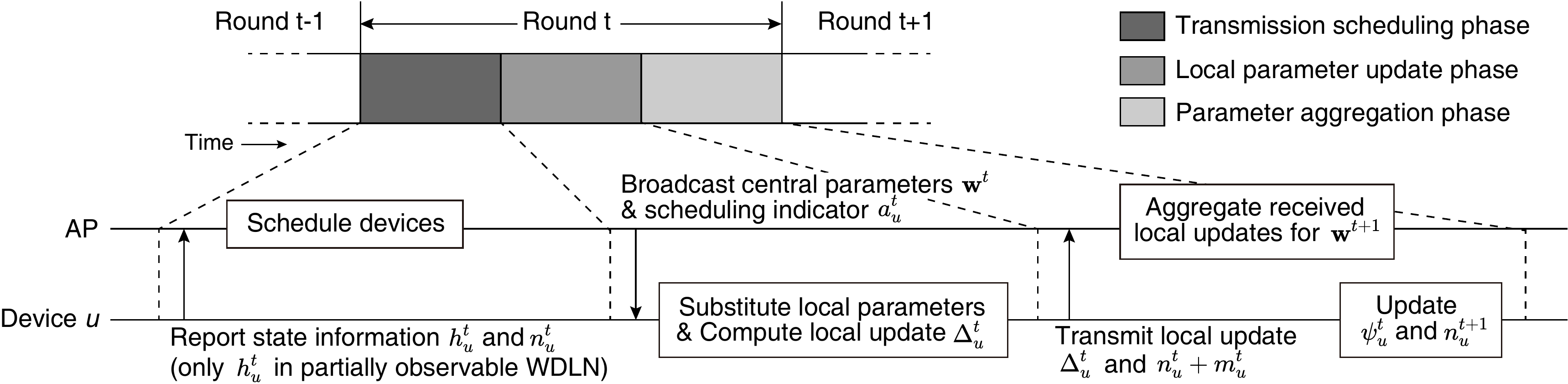}
	\caption{\label{fig:asynchFL}The asynchronous FL procedure in the WDLN.}
\end{figure*}

First, the AP schedules the devices to transmit their local updates based on the state information to effectively utilize the limited radio resources in each round.
We define the maximum number of devices that can be scheduled in each round as $W$.
It is worth noting that $W$ is restricted by the bandwidth of the WDLN and typically much smaller than the number of devices $U$ (i.e., $W\ll U$).
We define a transmission scheduling indicator of device $u$ in round $t$ as $a_u^t\in\{0,1\}$, where 1 represents device $u$ is scheduled to transmit its gradient in round $t$ and 0 represents it is not.
The vector of the transmission scheduling indicators in round $t$ is defined as $\ba^t=\{a_u^t\}_{u\in\calU}$.
In round $t$, the AP schedules the devices to transmit their local updates satisfying the following constraint:\looseness=-1
\begin{equation}
\sum_{u\in\calU} a^t_u=W.
\end{equation}
We then define the successful transmission indicator of the local update of device $u$ in round $t$ as a Bernoulli random variable $x_u^t$, where 1 represents the successful transmission and 0 represents the transmission failure.
Since the probability of the successful transmission of device $u$ in round $t$ is determined according to the channel gain of device $u$,
we can model it by using the approximation of the packet error rate (PER) with the given signal-to-interference-noise ratio (SINR) \cite{xi2011general,ferrand2015approximations}.
Moreover, we can use the PER provided in \cite{wu2013energy,ge2014packet} to model it considering the HARQ or ARQ retransmissions.
For example, the PER of an uncoded packet can be given by
$\mathbb{P}[x=0|\Phi]=1-(1-b(\Phi))^n$, where $\Phi$ is the SINR with the channel gain $h$, $b(\cdot)$ is the bit error rate for the given SINR, and $n$ is the packet length in bits.
We refer the readers to \cite{ferrand2015approximations} for more examples of the PER approximations with coding.
The vector of the successful transmission indicators in round $t$ is defined as $\bx^t=\{x_u^t\}_{u\in\calU}$.

We assume the number of the samples that have arrived at device $u$ during round $t$, $m_u^t$, is an independently and identically distributed (i.i.d.) Poisson random variable with a system parameter $\theta_u$.\footnote{It is worth noting that the assumption of Poisson data arrivals is only for simple implementation of the Bayesian approach in our proposed method. We can easily generalize our proposed method for any other i.i.d. data arrivals by using sampling algorithms such as Gibbs sampling.}
At the end of round $t$ (i.e., at the beginning of round $t+1$), the number of aggregated samples of device $u$ from the latest successful local update transmission to the beginning of round $t+1$,\footnote{For brevity, to denote $n_u^{t+1}$, we simply write ``the number of aggregated samples of device $u$ at the beginning of round $t+1$'' in the rest of this paper.} $n_u^{t+1}$, and the total number of samples from device $u$ used for the central updates before the beginning of round $t+1$, $N_u^{t+1}$, should be calculated depending on the local update transmission as follows:
\begin{equation}
\label{eqn:aggregated_samples}
n_u^{t+1}=\begin{cases} 0,& a_u^t=1\textrm{ and }x_u^t=1 \\
n_u^t+m_u^t,&\textrm{otherwise}\end{cases}
\end{equation}
and
\begin{equation}
\label{eqn:total_sample}
N_u^{t+1}=\begin{cases} N_u^t+n_u^t+m_u^t,& a_u^t=1\textrm{ and }x_u^t=1 \\
N_u^t,&\textrm{otherwise.}\end{cases}
\end{equation}
With the scheduling and successful transmission indicators, $a_u^t$'s and $x_u^t$'s, we can rewrite the equation for the centralized parameter update at the AP in \eqref{eqn:parameter_update} as follows:
\begin{equation}
\label{eqn:updating_parameters}
\bw^{t+1}=\bw^t-\sum_{u\in\mathcal{U}}a_u^t x_u^t c_u^t\Delta_u^t.
\end{equation}
We summarize the asynchronous FL procedure in the WDLN in Algorithm \ref{alg:asynchFL} and illustrate in Fig. \ref{fig:asynchFL}.

\algnewcommand{\LineComment}[1]{\Statex \(\triangleright\) #1}
\setlength{\textfloatsep}{5pt}
\begin{algorithm}[t]
	\caption{Asynchronous FL in WDLN}
	\label{alg:asynchFL}
	\begin{algorithmic}[1]
		\State \textbf{Input}: Regularization parameter $\lambda$, decay coefficient $\beta$
		\State Initialize variables $N_u=n_u=0$
		\For{$t=1,2,...$}
		\LineComment \textit{Transmission scheduling phase}
		\parState{The AP observes $\bh^t$ and $\bn^t$ and schedules the update transmissions $\ba^t$}
		\LineComment \textit{Local parameter update phase}
		\parState{The AP broadcasts $\bw^t$ and devices updates its local parameters}
		\State Each device computes its local update as in \eqref{eqn:local_gradient_aggregating},\eqref{eqn:local_update},\eqref{eqn:dynamic_learning_rate}
		\LineComment \textit{Parameter aggregation phase}
		\parState{Each scheduled device transmits its local update $\Delta_u^t$ and $n_u^t+m_u^t$ to allow the AP to calculate $c_u^t$}\label{alg:line:transmission}
		\parState{The AP updates the central parameters as in \eqref{eqn:updating_parameters} and $N_u^{t+1}$ as in \eqref{eqn:total_sample}}
		\State Each device updates $n_u^{t+1}$ as in \eqref{eqn:aggregated_samples} and $\psi_u^t$ as in \eqref{eqn:gradient_balancing}
		\EndFor
	\end{algorithmic}
\end{algorithm}

\subsection{Convergence Analysis of Asynchronous FL in WDLN}

We analyze the convergence of the asynchronous FL procedure proposed in the previous subsection.
We first introduce some definitions and an assumption on the objective function of asynchronous FL in \eqref{eqn:problem} for the analysis.

\begin{definition}
($L$-smoothness)
The function $f$ is $L$-smooth if it has Lipschitz continuous gradient with constant $L>0$ (i.e., $\forall \bx_1, \bx_2$),
$
f(\bx_1)-f(\bx_2)\leq \left<\nabla f(\bx_2),\bx_1-\bx_2\right>+\frac{L}{2}||\bx_1-\bx_2||^2.
$
\end{definition}
\begin{definition} ($\xi$-strongly convexity)
The function $f$ is $\xi$-strongly convex with $\xi>0$ if $\forall \bx_1, \bx_2$,
$
f(\bx_1)-f(\bx_2)\geq \left<\nabla f(\bx_2),\bx_1-\bx_2\right>+\frac{\xi}{2}||\bx_1-\bx_2||^2.
$
\end{definition}
\begin{definition}
(Bounded gradient dissimilarity) The local functions are $V$-locally dissimilar at $\bw$ if $\mathbb{E}[||(\psi_u)||^2]\leq ||\nabla l(\bw)||^2V^2$, where $\psi_u$ is the aggregated local gradients of device $u$ in \eqref{eqn:gradient_balancing} and $l(\bw)$ is the central loss function in \eqref{eqn:problem}.
\end{definition}

\begin{assumption}
\label{assumption1}
The objective function of asynchronous FL, $l(\bw)$, in \eqref{eqn:problem} is bounded from below, and there exists $\epsilon>0$ such that 
$\nabla l(\bw)^\top \mathbb{E}[\psi_u]\geq \epsilon ||\nabla l(\bw)||^2$ hold for all $\bw$.\looseness=-1
\end{assumption}
It is worth noting that this assumption is a typical one used in literature \cite{MLSYS2020_38af8613,xie2019asynchronous,chen2019asynchronous}.
With this assumption, we can show the convergence of the asynchronous FL procedure in the WDLN in the following theorem.

\begin{theorem}
\label{thm:convergence}
Suppose that the objective function of asynchronous FL in the WDLN, $l(\bw)$, in \eqref{eqn:problem} is $L$-smooth and $\xi$-strongly convex, and the local gradients $\psi_u$'s are $V$-dissimilar.
Then, if Assumption \ref{assumption1} holds, after $T$ rounds, the asynchronous FL procedure in the WDLN satisfies
\begin{IEEEeqnarray}{l}
\mathbb{E}[l(\bw^T)-l(\bw^*)]\leq \label{eqn:thm1_eq1} \\
\prod_{t=0}^{T-1}\left\lbrace 1\!+\! 2\xi\bar{\eta}^t \left(\frac{L\bar{\eta}^tW^2V^2}{2}\!-\! \epsilon\!\!\sum_{u\in\calU}\!a_u^t \mathbb{P}[x_u^t\!=\!1]\right)\right\rbrace(l(\bw^0)\!-\!l(\bw^*)). \nonumber
\end{IEEEeqnarray}
In addition, if there exists any $\zeta >0$ satisfies $\sum_{u\in\calU}a_u^t\mathbb{P}[x_u^t=1]\geq \zeta$ and $\underline{\eta}\leq \eta_u^t < \eta^t=\frac{2\epsilon\zeta}{LW^2V^2}(\max_{u\in\calU}\{c_u^t\})^{-1}$, the following holds:
\begin{equation}
\mathbb{E}[l(\bw^T)-l(\bw^*)]\leq(1-2\xi\underline{\eta} \epsilon')^T(l(\bw^0)-l(\bw^*)), \label{eqn:thm1_eq2}
\end{equation}
where $\epsilon'=\epsilon\zeta-\frac{L\bar{\eta} W^2V^2}{2}$ and $\bar{\eta}=\max_{\forall t}\eta^t$.
\end{theorem}
\begin{IEEEproof}
See Appendix \ref{appendix:thm1}.
\end{IEEEproof}
Theorem \ref{thm:convergence} takes consideration of the asynchronous FL procedure in the WDLN compared with the convergence analysis of asynchronous FL in \cite{chen2019asynchronous}.
As a result, the theorem implies that the faster convergence can be expected as a larger expected number of successful transmissions in the WDLN, $\sum_{u\in\calU}a_u^t\mathbb{P}[x_u^t=1]$.
Besides, it theoretically shows the convergence of the asynchronous FL procedure in the WDLN as $T\rightarrow\infty$ if the expected number of successful transmissions is non-zero since $0<2\xi\underline{\eta} \epsilon'<1$.
However, various aspects in learning, such as a learning speed and a robustness to stragglers, are not directly shown in this theorem, and they may appear differently according to transmission scheduling strategies used in the asynchronous FL procedure as will be shown in Section \ref{sec:experiments}.

\section{Problem Formulation}
\label{sec:problem_formulation}
We now formulate an asynchronous learning-aware transmission scheduling (ALS) problem to maximize the performance of FL learning.
In the literature, it is empirically shown that the FL learning performance mainly depends on the local update of each device, which is determined by the arrived data samples at the devices \cite{lim2020federated,xie2019asynchronous}.
However, it is still not clearly investigated which characteristics in a local update bring a larger impact than others to learn the optimal parameters \cite{216799,lim2020federated}.
Moreover, it is hard for the AP to exploit information relevant to the local updates because the AP cannot obtain the local updates before the devices transmit them to the AP and cannot access the data samples of the devices due to the limited radio resources and privacy issues.
Due to these reasons, instead of directly finding the impactful local updates, the existing works \cite{chen2020joint,amiriconvergence,shi2020joint} schedule the transmissions according to the factors in the bound of the convergence rate of FL, which implicitly represent the impact of the local updates in terms of the convergence.
However, as pointed out in Section \ref{sec:introduction}, the existing works on transmission scheduling for FL do not consider asynchronous FL over multiple rounds.
As a result, their scheduling criterion based on the convergence rate may become inefficient in asynchronous FL since in a long-term aspect with asynchronous FL, the deterioration of the central model's convergence due to the absence of some devices in model aggregation can be compensated later thanks to the nature of asynchronous FL.

Contrary to the existing works, we focus on maximizing the average number of learning from the local data samples considering asynchronous FL.
The number of data samples used in learning can implicitly represent the amount of learning.
Roughly speaking, a larger number of data samples leads the empirical loss in \eqref{eqn:problem} to be converged into a true one in the real-world.
It is also empirically accepted in the literature that more amount of data samples generally results in better performance \cite{domingos2012few,goodfellow2016deep}.
In this context, in FL, the number of the local samples used to calculate the local gradients represents the amount of learning since the central parameters are updated by aggregating the local gradients.
However, in asynchronous FL, the delayed local gradients due to scheduling or transmission failure are aggregated to the current one, which may cause an adverse effect on learning.
Hence, here, we maximize the total number of samples used to calculate the local gradients aggregated in the central updates while considering their delay as a cost to minimize its adverse effect on learning.

We define a state in round $t$ by a tuple of the channel gain $\bh^t$ and the numbers of the aggregated samples $\bn^t$, $\bs^t=(\bh^t,\bn^t)$ and the state space $\calS\subset\mathbb{R}^{U}\times\mathbb{N}^U$.
The AP can observe the state information in the transmission scheduling phase of each round.
It is worth noting that in the following section, we will also consider the more restrictive environments in which the AP can observe only a \textit{partial} state in round $t$ (i.e., the channel gain $\bh^t$).
We denote system uncertainties in the WDLN, such as the successful local update transmissions and data sample arrivals, by random disturbances. 
The vector of the random disturbances in round $t$ is defined as $\bv^t=(\bx^t,\bm^t)$, where $\bx^t$ is the vector of the successful transmission indicators in round $t$ and $\bm^t$ is the vector of the number of samples that have arrived at each device during round $t$.
Then, an action in round $t$, $\ba^t$, is chosen from the action space, defined by $\calA=\{\ba:\sum_{u\in\calU}a_u=W\}$.
In each round $t$, the AP determines the action based on the observed state while considering the random disturbances which are unknown to the AP.
To evaluate the effectiveness of the transmission on learning in each round, we define a metric, called an \textit{effectivity score} of asynchronous FL, as\looseness=-1
\begin{equation}
\label{eqn:objective_function}
F(\bs^t,\ba^t,\bv^t)=\sum_{u\in\bar{\calU}^t} n_u^t+ m_u^t - \sum_{u\in\calU\setminus\bar{\calU}^t} \gamma(n_u^t+m_u^t),
\end{equation}
where $\bar{\calU}^t$ is the set of the devices whose local update transmission was successful in round $t$ and $\gamma\in[0,1)$ is a delay cost coefficient to consider the adverse effect of the stragglers.
Note that this metric is correlated over time since $n_u^t$ depends on the previous scheduling and the transmission result.
The first term of the effectivity score represents the total number of the samples that will be effectively used in the central update while the second term represents the total number of the samples that may cause the adverse effects in learning due to the delayed local updates.
We now formulate the ALS problem maximizing the total expected effectivity score of asynchronous FL as
\begin{IEEEeqnarray}{cl}
\label{eqn:problem_ALS}
\maximize_\pi &~\lim_{T\rightarrow\infty}\mathbb{E}\left[\sum_{t=1}^T F(\bs^t,\ba^t,\bv^t)\right],
\end{IEEEeqnarray}
where $\pi:\mathcal{S}\rightarrow\mathcal{A}$ is a policy that maps a state to an action.
This problem maximizes the number of samples used in asynchronous FL while trying to minimize the adverse effect of the stragglers due to the delay.

\begin{remark}
We highlight that the proposed effective score for asynchronous FL shares the same principle with the metrics in the conventional works \cite{chen2020joint,shi2020joint}.
Roughly speaking, the metrics try to maximize the number of data samples used in learning the transmitted local updates since it maximizes the convergence rates of the central model according to the convergence analysis of synchronous FL.
Similar to this, the proposed effective score tries to maximize the number of data samples used in learning the successfully transmitted local updates to maximize the amount of learning conveyed to the AP.
However, at the same time, it tries to minimize the number of data samples used in learning the delayed local updates to minimize their adverse effects, while the conventional metrics do not introduce any penalty due to the delayed local updates.
\end{remark}

We define the system parameters related to the system uncertainties as 
\begin{equation}
\label{eqn:parameters}
\btheta=\{\btheta_1^C,...,\btheta_U^C,\theta_1^P,...,\theta_U^P\}\in\Theta,
\end{equation}
where $\Theta$ is a system parameter space.
These system parameters express the system uncertainties, such as the channel statistics and the average number of arrived samples (i.e., arrival rate) of each device in each round, in the WDLN, and thus, we can solve the problem if they are known.
We define the true system parameters for the WDLN as $\btheta^*=\{\btheta^{*,C},\btheta^{*,P}\}$, where $\btheta^{*,C}$ and $\btheta^{*,P}$ are the true system parameters for the channel gains and the arrival rates, respectively.
Formally, if the system parameters $\btheta^*$ are perfectly known in advance as a priori information, the state transition probabilities, $\mathbb{P}[\bs'|\bs,\ba]$, can be derived by using the probability of successful transmission, the distribution of the channel gains, and the Poisson distribution with the a priori information.
Then, based on the transition probabilities, the problem in \eqref{eqn:problem_ALS} can be optimally solved by standard dynamic programming (DP) methods \cite{bertsekas1995dynamic}.
However, this is impractical since it is hard to obtain such a priori information in advance.
Besides, even if such a priori information is perfectly known, the computational complexity of the standard DP methods is too high as will be shown in Section \ref{sec:complexity}.
Hence, we need a learning approach such as reinforcement learning (RL) that learns the information while proceeding the asynchronous FL, and develop algorithms based on it in the following section.

\section{Asynchronous Learning-Aware Transmission Scheduling}

In this section, we develop three ALS algorithms each of which requires different information to solve the ALS problem:
an ALS algorithm with perfect a priori information (ALSA-PI), a Bayesian ALS algorithm (BALSA), and a Bayesian ALS algorithm with a partially observable state information (BALSA-PO).
We summarize the information required in ASLA-PI, BALSA, and BALSA-PO in Table \ref{table:used_information}.
In the table, we can see that from ALSA-PI to BALSA-PO, the required information in those three algorithms is gradually reduced.

\begin{table}[!t]
	\fontsize{8pt}{1em}\selectfont
	\centering
	\caption{Summary of information required in each algorithm}
	\label{table:used_information}
	\begin{tabular}{c|c|c|c}
		\hline
		& \textbf{ALSA-PI} & \textbf{BALSA} & \textbf{BALSA-PO}\\\hline
		\makecell{\textbf{a priori information on}\\\textbf{true system parameters}} & $\btheta^{*,P}$ & - & - \\\hline
		\makecell{\textbf{State information}\\\textbf{reported from the devices}} & $\bh,\bn$ & $\bh,\bn$ & $\bh$ \\\hline
	\end{tabular}
\end{table}

\subsection{Parametrized Markov Decision Process}
Before developing algorithms, we first define a \textit{parameterized} Markov decision process (MDP) based on the ALS problem in the previous section.
Since the system uncertainties in the WDLN, such as the channel statistics and the arrival rates of the samples, depend on the system parameters $\btheta$ in \eqref{eqn:parameters}, the MDP is parameterized by system parameters $\btheta$.
In specific, it is defined as $M_\btheta=(\calS,\calA,r^\btheta,P^\btheta)$, where $\calS$ and $\calA$ are the state space and action space of the ALS problem, respectively, $r^\btheta$ is the reward function, and $P^\btheta$ is the transition probability such that $P^\btheta(\bs'|\bs,\ba)=\mathbb{P}(\bs^{t+1}=\bs'|\bs^t=\bs,\ba^t=\ba,\btheta)$.
We define the reward function $r^\btheta$ as the effectivity score in \eqref{eqn:objective_function}.
For theoretical results, we assume that the state space $\calS$ is finite and the reward function is bounded as $r^\btheta:\calS\times\calA\rightarrow[0,1]$.\footnote{We can easily implement such a system by quantizing the channel gain and truncating $n_u^t$ and $m_u^t$ with a large constant. Then, the state space becomes finite and we can use the reward function as the normalized version of $F(\bs,\ba)$ with the maximum expected reward that can be derived from the constant. In practice, this system is more realistic to be implemented since such variables in a real system are typically finite.\looseness=-1}
The average reward per round of a stationary policy $\pi$ is defined as
\begin{equation}
J_\pi(\btheta)=\lim_{T\rightarrow\infty}\frac{1}{T}\mathbb{E}\Big[\sum_{t=1}^T r^\btheta(\bs^t,\ba^t)\Big].
\end{equation}
The optimal average reward per round $J^*(\btheta)=\max_\pi J_\pi(\btheta)$ satisfies the Bellman equation:
\begin{equation}
\label{eqn:bellman_equation}
J^*(\btheta)+v(\bs,\btheta)\!=\max_{a\in\mathcal{A}}\Big\{r^\btheta(\bs,\ba)+\!\sum_{\bs'\in\calS}\!P^\btheta(\bs'|\bs,\ba)v(\bs'\!,\!\btheta)\Big\},~\forall	\bs\!\in\!\calS,
\end{equation}
where $v(\bs,\btheta)$ is the value function at state $\bs$.
We can define the corresponding optimal policy, $\pi^*(\bs,\btheta)$, as the maximizer of the above optimization.
Then, the policy $\pi^*(\bs,\btheta^*)$ becomes the optimal solution to the problem in \eqref{eqn:problem_ALS}.

\subsection{ALSA-PI: Optimal ALS with Perfect a Priori Information}
\label{sec:alsa-pi}
We now develop ALSA-PI to solve the ALS problem when $\btheta^*$ is perfectly known at the AP in advance.
Typically, the true system parameters $\btheta^*$ is unknown, but here we introduce ALSA-PI since it will be used to develop BALSA in Section \ref{sec:balts}. Besides, in some cases, $\btheta^*$ might be precisely estimated by using past experiences.
Since the true system parameters $\btheta^*$ are known, the problem in \eqref{eqn:problem_ALS} can be solved by finding the optimal policy to $\btheta^*$.
However, in general, computing the optimal policy $\pi^*$ to given parameters requires a large computational complexity which exponentially increases with the number of devices, which is often called a curse of dimensionality.
Hence, even if the true system parameters $\btheta^*$ are perfectly known in advance, it is hard to compute the corresponding optimal policy $\pi^*(\bs,\btheta^*)$.
However, for the ALS problem, a greedy policy, which myopically chooses the action to maximize the expected reward in the current round, becomes the optimal policy as the following theorem.
\begin{theorem}
\label{thm:greedy_optimal}
For the parameterized MDP with given finite parameters, the greedy policy in \eqref{eqn:greedy_policy} is optimal.
\end{theorem}
\begin{IEEEproof}
See Appendix \ref{appendix:thm2}.
\end{IEEEproof}
With this theorem, we can easily develop ALSA-PI by adopting the greedy policy with the known system parameters $\btheta^*$ thanks to the structure of the reward.
This also implies that we can significantly reduce the computational complexity to solve the ALS problem with the known parameters compared with the DP methods that are typical ones to solve MDPs.
The computational complexity will be analyzed in Section \ref{sec:complexity}.

In round $t$, the greedy policy for given system parameters $\btheta$ chooses the action by solving the following optimization problem:
\begin{equation}
\maximize_{\ba\in\calA} \mathbb{E}[F(\bs^t,\ba,\bv^t)|\btheta],
\end{equation}
where $\mathbb{E}[\cdot|\btheta]$ represents the expectation over the probability distribution of the given system parameters $\btheta$.
With the scheduling and successful transmission indicators, $a_u^t$'s and $x_u^t$'s, we rearrange the objective function in \eqref{eqn:objective_function} as
\begin{equation}
\label{eqn:objective_function2}
F(\bs^t,\ba,\bv^t)=\sum_{u\in\calU}a_ux_u^t(1+\gamma)(n_u^t+ m_u^t)-\gamma(n_u^t+ m_u^t).
\end{equation}
Then, we can reformulate the problem as
\begin{equation}
\maximize_{\ba\in\calA} \sum_{u\in\calU}a_u\mathbb{E}[ x_u^t(n_u^t+m_u^t)|\btheta^P],
\end{equation}
since the last term in \eqref{eqn:objective_function2} does not depend on both transmission scheduling indicator $\ba$ and uncertainty on the successful transmission $\bx^t$.
In addition, the conditional expectation in the problem depends only on the system parameters for the arrival rates of the data samples, $\btheta^P=\{\theta_1^P,...\theta_U^P\}$, because $x_u^t$ is solely determined by the current channel gain.
To solve the problem, the AP estimates the expected number of samples of each device $u$, $\mathbb{E}[x_u^t(n_u^t+m_u^t)|\btheta^P]$, by using its channel gain $h_u^t$ and the system parameter $\theta_u^P$.
Let the devices be sorted in descending order of their expected number of samples $\mathbb{E}[x_u^t(n_u^t+m_u^t)|\btheta^P]$ and indexed by $(1),(2),...,(U)$.
Then, the greedy policy in round $t$ is easily obtained as
\begin{equation}
\label{eqn:greedy_policy}
\pi^g(\bs^t,\btheta^P)=\begin{cases} a_u=1, & u\in\{(1),...,(W)\} \\ a_u=0,&\textrm{otherwise.}\end{cases}
\end{equation}
In ALSA-PI, the AP schedules the devices according to the greedy policy with the true system parameter $\btheta^{*,P}$ (i.e., $\pi^g$ in \eqref{eqn:greedy_policy} with $\btheta^{*,P}$) in each round.

\begin{remark}
For the time-varying maximum number of devices that can be scheduled in each round, the greedy policy in \eqref{eqn:greedy_policy} can be easily extended by substituting $W$ into $W^t$, where $W^t$ denotes the maximum number in round $t$.
This implies that the ALS algorithms (i.e., ALSA-PI, BALSA, and BALSA-PO) can be easily extended for it as well since they are implemented by using the greedy policy and in the methods, $W$ is related to the greedy policy only.
\end{remark}

\subsection{BALSA: Optimal Bayesian ALS without a Priori Information}
\label{sec:balts}

We now develop BALSA that solves the parameterized MDP without requiring any a priori information.
To this end, it adopts a Bayesian approach to learn the unknown system parameters $\btheta$.
We define the policy determined by BALSA as $\phi=(\phi^1,\phi^2,...)$ each of which $\phi^t(\bg^t)$ chooses an action according to the history of states, actions, and rewards, $\bg^t=(\bs^1,...,\bs^t,\ba^1,...,\ba^{t-1},r^1,...,r^{t-1})$.
To apply the Bayesian approach, we assume that the system parameters that belong to $\btheta$ are independent to each other.
We denote a prior distribution for the system parameters $\btheta$ by $\mu^1$.
For the prior distribution, the non-informative prior can be typically used if there is no prior information about the system parameters. On the other hand, if there is any prior information, then it can be defined by considering the information. For example, if we know that a parameter belongs to a certain interval, then the prior distribution can be defined as a truncated distribution whose probability measure outside of the interval is zero.
We then define the Bayesian regret of BALSA up to time $T$ as
\begin{equation}
R(T)=\mathbb{E}\Big[TJ^*(\btheta)-\sum_{t=1}^T r^{\btheta}(\bs^t,\phi^t(\bg^t))\Big],
\end{equation}
where the expectation in the above equation is over the prior distribution $\mu^1$ and the randomness in state transitions.
This Bayesian regret has been widely used in the literature of Bayesian RL as a metric to quantify the performance of a learning algorithm \cite{gopalan2015thompson,osband2017posterior,ouyang2017learning}.

In BALSA, the system uncertainties are estimated by a Bayesian approach.
Formally, in the Bayesian approach, the posterior distribution of $\btheta$ in round $t$, which is denoted by $\mu^t$, is updated by using the observed information about $\btheta$ according to the Bayes' rule.
Then, the posterior distribution is used to estimate or sample the system parameters for the algorithm.
We will describe how the posterior distribution is updated in detail later.
We define the number of visits to any state-action pair $(\bs,\ba)$ before time $t$ as
\begin{equation}
M^t(\bs,\ba)=|\{\tau<t:(\bs^\tau,\ba^\tau)=(\bs,\ba)\}|,
\end{equation}
where $|\cdot|$ denotes the cardinality of the set.

BALSA operates in stages each of which is composed of multiple rounds.
We denote the start time of stage $k$ of  BALSA by $t_k$ and the length of stage $k$ by $T_k=t_{k+1}-t_k$. With the convention, we set $T_0=1$.
Stage $k$ ends and the next stage starts if $t>t_k+T_{k-1}$ or $M^t(\bs,\ba)> 2M^{t_k}(\bs,\ba)$ for some $(\bs,\ba)\in\calS\times\calA$.
This balances the trade-off between exploration and exploitation in BALSA.
Thus, the start time of stage $k+1$ is given by
\begin{IEEEeqnarray}{l}
t_{k+1}=\min\{t>t_k:t>t_k+T_{k-1}\textrm{ or } \nonumber \\
\qquad M^t(\bs,\ba)> 2M^{t_k}(\bs,\ba)\textrm{ for some }(\bs,\ba)\in\calS\times\calA\},\quad
\end{IEEEeqnarray}
and $t_1=1$.
This stopping criterion allows us to bound the number of stages over $T$ rounds.
At the beginning of stage $k$, system parameters $\btheta_k$ are sampled from the posterior distribution $\mu^{t_k}$.
Then, the action is chosen by the optimal policy corresponding to the sampled system parameter $\btheta_k$, $\pi^*(\bs,\btheta_k)$, until the stage ends.
It is worth noting that this posterior sampling procedure, in which the system parameters sampled from the posterior distribution are used for choosing actions, has been widely applied to address the exploration-exploitation dilemma in RL \cite{gopalan2015thompson,osband2017posterior,ouyang2017learning,lee2021sdf}.
Since the greedy policy, $\pi^g(\bs,\btheta_k)$, is the optimal policy to the ALS problem as shown in the previous subsection, we can easily implement BALSA by using it.
Besides, this makes the AP not have to estimate the posterior distribution of the parameters for the channel gains, $\btheta^C=\{\btheta_1^C,...,\btheta_U^C\}$,
since the greedy policy do not use them.

Here, we describe BALSA in more detail.
In round $t$, the AP observes the states $\bs^t=(\bh^t,\bn^t)$.
Then, the AP can obtain the numbers of arrived samples of all devices during round $t-1$, $m_u^{t-1}$'s, as follows: for each device whose local model was transmitted to the AP in round $t-1$ (i.e., $u\in\bar{\calU}^{t-1}$), it can be known because $n_u^{t-1}+m_u^{t-1}$ is transmitted to the AP in round $t-1$ for the local update, and for the other devices (i.e., $u\in\calU\setminus\bar{\calU}^{t-1}$), it can be obtained by subtracting $n_u^{t-1}$ from $n_u^t$.
Then, the AP updates the posterior distribution of $\btheta^P$, $\mu^t_P$, according to Bayes' rule.
We denote the posterior distribution for a system parameter $\theta$ that belongs to $\btheta$ in round $t$ by $\mu^t_{\theta}$.
Then, the posterior distribution of $\btheta^P$ is updated by using $m_u$'s.
For example, if the non-informative Jeffreys prior for the Poisson distribution is used for the system parameter $\theta_u^P$, then its posterior distribution in round $t$ is derived as follows \cite{misgiyati2017bayesian}:
\begin{equation}
	\label{eqn:posterior_poisson}
	\mu^t_{\theta_u^P}(\theta|\bg^t)=\frac{(t-1)^{S+a}}{\Gamma(S+1/2)}\theta^{S-1/2}e^{-\theta(t-1)},
\end{equation}
where $S=\sum_{\tau=1}^{t-1}m_u^\tau$.
In this paper, we implement BALSA based on the non-informative prior.
In each stage $k$ of the algorithm, the AP samples the system parameters $\btheta^P_k$ by using the posterior distribution in \eqref{eqn:posterior_poisson} and schedules the local update transmission according to the greedy policy in \eqref{eqn:greedy_policy} using the sampled system parameters $\btheta^P_k$.
We summarize BALSA in Algorithm \ref{alg:balts_optimal}.\looseness=-1

\setlength{\textfloatsep}{5pt}
\begin{algorithm}[t]
	\caption{BALSA}
	\label{alg:balts_optimal}
	\begin{algorithmic}[1]
		\State \textbf{Input}: Prior distribution $\mu^1$
		\State Initialize $k\leftarrow 1$, $t\leftarrow 1$, $t_k\leftarrow 0$
		\While{TRUE}
		\State $T_{k-1}\leftarrow t-t_k$ and $t_k\leftarrow t$
		\State Sample $\btheta_k^P\sim\mu^{t_k}_P$
		\While{$t\!\leq\! t_k\!+\!T_{k-1}$ and $M^t(\bs,\ba)\leq 2M^{t_k}(\bs,\ba),~\forall(\bs,\ba)$}
		\State Choose action $\ba^t\leftarrow \pi^g(\bs^t,\btheta_k^P)$
		\State Observe state $\bs^{t+1}$ and reward $r^{t+1}$
		\parState{Update $\mu^{t+1}_P$ as in \eqref{eqn:posterior_poisson} using $m_u$'s for all devices}\label{alg:line:posterior}
		\State $t\leftarrow t+1$
		\EndWhile
		\State $k\leftarrow k+1$
		\EndWhile
	\end{algorithmic}
\end{algorithm}

We can prove the regret bound of BALSA as the following theorem.
\begin{theorem}
\label{thm:regret_bound}
Suppose that the maximum value function over the state space is bounded, i.e., $\max_{s\in\calS}v(\bs,\btheta) \leq H$ for all $\btheta\in\Theta$. Then, the Bayes regret of BALSA satisfies
\begin{equation}
R(T)\leq (H+1)\sqrt{2SAT\log T}+49HS\sqrt{AT\log AT},
\end{equation}
where $S$ and $A$ denote the numbers of states and actions, i.e., $|\calS|$ and $|\calA|$, respectively.
\end{theorem}
\begin{IEEEproof}
See Appendix \ref{appendix:thm3}.
\end{IEEEproof}
The regret bound in Theorem \ref{thm:regret_bound} is sublinear in $T$.
Thus, in theory, it is guaranteed that the average reward of BALSA per round converges to that of the optimal policy (i.e., $\lim_{T\rightarrow\infty} R(T)/T=0$), which implies the optimality of BALSA in terms of the long-term average reward per round.

\subsection{BALSA-PO: Bayesian ALS for Partially Observable State Information}
\label{sec:balts-po}
In the WDLN, the state, $\bs=(\bh,\bn)$, can be reported to the AP from the devices.
In typical wireless networks, reporting $n_u^t$'s to the AP needs data transmissions of the devices, which require exchanging control messages between the AP and the devices individually.
Besides, it is expected that the number of devices participating in FL is typically much larger than the capacity of wireless networks that represents the number of devices that can transmit the local updates simultaneously \cite{lim2020federated}.
Hence, it might be impractical (at least inefficient) that all individual devices report their numbers of the aggregated samples $\bn$ to the AP in each round because of the excessive exchanges of the control messages between the AP and the individual devices for reporting $\bn$.
On the other hand, the channel gains may not require such data transmissions in an uplink case since the AP can estimate the channel gains of the devices by using the pre-defined reference signals transmitted from the devices \cite{hou2011demodulation}.
In this context, in this section, we consider a WDLN in which in each round, the AP observes the channel gain $\bh$ only and the numbers of the aggregated samples $\bn$ are not reported to the AP.
We can model this as a partially observable MDP (POMDP) based on the description in Section \ref{sec:problem_formulation}, but the POMDP is often hard to solve because of the intractable computation.
In this section, we develop an algorithm to solve the ALS problem in the partially observable WDLN by slightly modifying BALSA, which is called BALSA-PO.

In the fully observable WDLN, the AP observes the numbers of the aggregated samples, $n_u^t$'s, from all devices in each round $t$.
In BALSA, $n_u^t$'s are used to choose the action according to the greedy policy in \eqref{eqn:greedy_policy} and to obtain the numbers of the arrived samples of all devices during round $t-1$, $m_u^{t-1}$'s, for updating the posterior distribution of $\theta_u^P$.
Hence, in the partially observable WDLN, where $n_u^t$'s are not observable,
the AP cannot choose the action according to the greedy policy as well as obtain $m_u^{t-1}$'s.
To address these issues, first, in BALSA-PO, the AP approximates $n_u^t$ by using the sampled system parameter of $\theta_u^P$ as $\tilde{n}_u^t=(T_u^n-1)\theta_{u,k}^P$, where $T_u^n$ is the number of the rounds from the latest successful local update transmission to the current round (i.e., $T_u^n=t-\max\{\tau<t:x_u^\tau=1 \textrm{ and } a_u^\tau = 1\}$) and $\theta_{u,k}^P$ is the sampled system parameter of $\theta_u^P$ in stage $k$.
Then, the AP can choose the action according to the greedy policy by using the approximated $n_u^t$'s, i.e., $\tilde{n}_u^t$.
However, the chosen action will be meaningful only if the posterior distribution keeps updated correctly since $n_u^t$'s are approximated based on the sampled system parameters.
In the partially observable WDLN, it is a challenging issue updating the posterior distribution correctly since the AP cannot obtain $m_u^{t}$'s which are required to update the posterior distribution.
However, fortunately, even in the partially observable WDLN, the AP can still observe the information about the number of samples $n_u^t+m_u^t$ in every successful local update transmission since the information is included in the local update transmission (line \ref{alg:line:transmission} of Algorithm \ref{alg:asynchFL}).
Note that $n_u^t+m_u^t$ denotes the sum of $m_u^t$'s over the rounds from the latest successful local update transmission to the current round.
This sum of $m_u^t$'s in the partially observable WDLN is less informative than all individual values of $m_u^t$'s.
Nevertheless, it is informative enough to update the posterior distribution because the update of the posterior distribution of $\theta_u^P$ requires only the sum of $m_u^t$'s as in \eqref{eqn:posterior_poisson}.
Hence, in the partially observable WDLN, the AP can update the posterior distribution of $\theta_u^P$ when device $u$ successfully transmits its local update.
Then, BALSA-PO can be implemented by substituting the state in BALSA, $\bs$, to $\tilde{\bs}=(\bh,\tilde{\bn})$, where $\tilde{\bn}=\{\tilde{n}_1,...,\tilde{n}_U\}$ and changing the posterior distribution update procedure from BALSA in line \ref{alg:line:posterior} of Algorithm \ref{alg:balts_optimal} as follows: ``Update $\mu_{\theta_u^P}^{t+1}$ as in \eqref{eqn:posterior_poisson} using $n_u^t+m_u^t$ for device $u\in\bar{\calU}^t$''.

\begin{table}[!t]
	\fontsize{8pt}{1em}\selectfont
	\centering
	\caption{Comparison of computational complexity}
	\label{table:computational_complexity}
	\begin{tabular}{c|c}
		\hline
		& \textbf{Computational complexity up to round $T$}\\\hline
		\makecell{\textbf{DP-based with}\\\textbf{perfect information}} & $O(S^2A)$ \\\hline
		\makecell{\textbf{DP-based without}\\\textbf{information}} & \makecell[l]{Upper-bound: $O(S^2AT)$\\ Lower-bound: $O(\sqrt{S^3AT(\log T)^{-1}})$} \\\hline
		\textbf{ALSA-PI,BALSA(-PO)} & $O(UT\log U)$\\	\hline
		\multicolumn{2}{c}{$S=|\calS|=(|\mathcal{H}||\mathcal{N}|)^U$ and $A=|\calA|=\frac{U!}{W!(U-W)!}$} \\
		\hline
	\end{tabular}
\end{table}

\subsection{Comparison of Computational Complexity}
\label{sec:complexity}
We compare the computational complexities of the DP-based algorithms with/without a priori information, ALSA-PI, and BALSAs.
The DP-based algorithm with perfect a priori information requires to run one of the standard DP methods once to compute the optimal policy for the system parameters in the perfect information.
For the complexity analysis to compute the optimal policy, we consider the well-known value iteration whose computational complexity is given by $O(S^2A)$, where $S=|\calS|=(|\mathcal{H}||\mathcal{N}|)^U$, $A=|\calA|=\frac{U!}{W!(U-W)!}$, $\mathcal{H}$ is the set of the possible channel gain, and $\mathcal{N}$ is the set of the possible number of the aggregated samples.
Hence, the computational complexity of the DP-based algorithm with perfect a priori information is given by $O(S^2A)$.
The DP-based algorithm without a priori information denotes a learning method based on the posterior sampling in \cite{ouyang2017learning}. It operates over multiple stages as in BALSAs but the optimal policy for the sampled system parameters $\btheta_k$ in stage $k$ is computed based on the standard DP methods.
Thus, its upper-bound complexity up to round $T$ is given by $O(S^2AT)$ if the policy is computed in all rounds. Since the system parameters are sampled once for each stage, we can also derive the lower bound of its computational complexity up to round $T$ as $O(\sqrt{S^3AT(\log T)^{-1}})$ from Lemma \ref{lemma:stage_number} in Appendix \ref{appendix:thm3}.

Contrary to the DP-based algorithms, ALSA-PI and BALSAs use the greedy policy that finds the devices with the $W$ largest expected number of samples in each round. Thus, their computational complexity up to round $T$ is given by $O(UT\log U)$, where $U$ is the number of devices, according to the complexity of typical sorting algorithms.
It is worth emphasizing that the complexity of BALSAs grows according to the rate of $U\log U$ while the complexity of the DP algorithms exponentially increases according to $U$.
Besides, the DP-based algorithms have much larger computational complexities than BALSAs in terms of not only the asymptotic behavior but also the complexity coefficients in the Big-O notation since the DP-based algorithms require a complex computations based on the Bellman equation while BALSAs require only the computations to update posteriors in \eqref{eqn:posterior_poisson} and to sort the expected number of samples.
From the analyses, we can see that the complexity of BALSAs is significantly lower than that of the DP-based algorithms.
The computational complexities of the algorithms are summarized in Table \ref{table:computational_complexity}.\looseness=-1

\section{Experimental Results}
\label{sec:experiments}
In this section, we provide experimental results to evaluate the performance of our algorithms. To this end, we develop a dedicated Python-based simulator on which the following simulation and asynchronous FL are run.
We consider a WDLN composed of one AP and 25 devices.
We use a shard as a unit of data samples each of which consists of multiple data samples.
The number of data samples in each shard is determined according to datasets.
The arrival rate of the data samples of each device (i.e., the system parameter $\theta_u^P$), and the distance between the AP and each device are provided in Table \ref{table:devices_setting}.
The maximum number of scheduled devices in the WDLN, $W$, is set to be 5 as in \cite{xia2020multi}.
The channel gains are composed of the Rayleigh small-scale fading that follows an independent exponential distribution with unit mean and the large-scale fading based on the pathloss model $128.1+37.6\log_{10}(d)$, where the pathloss exponent is given by 3.76 and $d$ represents the distance in km.
The transmission power of each device is set to be 23 dBm and the noise power is set to be -96 dBm.
In the transmission of the local updates, the PER is approximated according to given SNR as in \cite{wu2013energy} based on turbo code.
The delay cost coefficient $\gamma$ in the ALS problem is set to be 0.01.
In asynchronous FL, we set the local learning rate $\eta$ to be 0.01 and the decay coefficient $\beta$ to be 0.001.\looseness=-1

\begin{table}[!t]
\footnotesize
\centering
\caption{Simulation Settings of Devices}
\label{table:devices_setting}
\setlength{\tabcolsep}{2pt}
\begin{tabular}{c|c|c|c|c|c|c|c|c}
\hline
\textbf{Device index} & \textbf{1-3} & \textbf{4-6} & \textbf{7-9} & \textbf{10-12} & \textbf{13-15} & \textbf{16} & \textbf{17} & \textbf{18}\\\hline
Arr. rate (shards/round) & 1 & 1 & 1 & 1 & 1 & 3 & 3 & 3 \\\hline
Distance (m) & 100 & 200 & 300 & 400 & 500 & 300 & 350 & 400 \\\hline\hline
\textbf{Device index} & \textbf{19} & \textbf{20} & \textbf{21} & \textbf{22} & \textbf{23} & \textbf{24} & \textbf{25} & \textbf{-} \\\hline
Arr. rate (shards/round) & 3 & 5 & 5 & 5 & 5 & 10 & 10 & - \\\hline
Distance (m) & 450 & 300 & 350 & 400 & 450 & 400 & 450 & - \\\hline
\end{tabular}
\end{table}

In the experiments, we consider MNIST
and CIFAR-10 datasets.
For the MNIST dataset, we set the number of data samples in each shard to be 10 and consider
a convolutional neural network model with two $5\times 5$ convolution layers with $2\times2$ max pooling, a fully connected layer, and a final softmax output layer.
The first convolution layer is with 1 channel and the second one is with 10 channels.
The fully connected layer is composed of 320 units with ReLU activation.
When training the local models for MNIST, we set the local batch size to be 10 and the number of local epochs to train the local model to be 10.
For the CIFAR-10 dataset, we set the number of data samples in each shard to be 50 and consider a well-known VGG19 model \cite{Simonyan15}. We also set the local batch size to be 50 and the number of local epochs to be 5.

To evaluate the performance of our algorithms, we compare them with an ideal benchmark and state-of-the-art scheduling algorithms.\footnote{In this paper, we do not compare our algorithms with model compressing methods \cite{chen2018lag,lin2018deep,xu2020ternary}, which reduces the communication costs of a single local model transmission, since they can be used with our algorithms orthogonally.}
The descriptions of the algorithms are provided as follows.
\begin{itemize}
\item \textbf{Bench} represents an ideal benchmark algorithm, where in each round, the AP updates the central parameters by aggregating the local updates of all devices as in \texttt{FedAvg} \cite{mcmahan2017communication}.
This provides the upper bound of the model performance because it is based on an ideal system with no radio resource constraints and transmission failure.
 
\item \textbf{RR} represents an algorithm in which the devices are scheduled in a round robin manner \cite{yang2019scheduling}. This algorithm does not consider both arrival rates of the data samples and channel information of the devices.

\item \textbf{$W$-max} represents an algorithm that schedules the devices who have the $W$ strongest channel gains. This algorithm does not consider the arrival rates of the data samples. It can represent the scheduling strategies in \cite{amiriconvergence,chen2020joint}.

\item \textbf{ALSA-PI} is implemented as described in Section \ref{sec:alsa-pi} to schedule the devices according to the greedy policy in \eqref{eqn:greedy_policy} with the perfect a priori information about $\btheta^{*,P}$.

\item \textbf{BALSA} is implemented as Algorithm \ref{alg:balts_optimal} in Section \ref{sec:balts} for the fully observable WDLN. For the Bayesian approach, the Jeffreys prior is used.

\item \textbf{BALSA-PO} is implemented as described in Section \ref{sec:balts-po}. The Jeffreys prior is used as in BALSA. This algorithm is for the partially observable WDLN.
\end{itemize}
The models are trained by asynchronous FL with above transmission scheduling algorithms.
We run 50 simulation instances for MNIST dataset and 20 simulation instances for CIFAR-10 dataset.
In the following figures, the 95\% confidence interval is illustrated as a shaded region.

\begin{figure}[!t]
	\centering
	\includegraphics[scale=0.65]{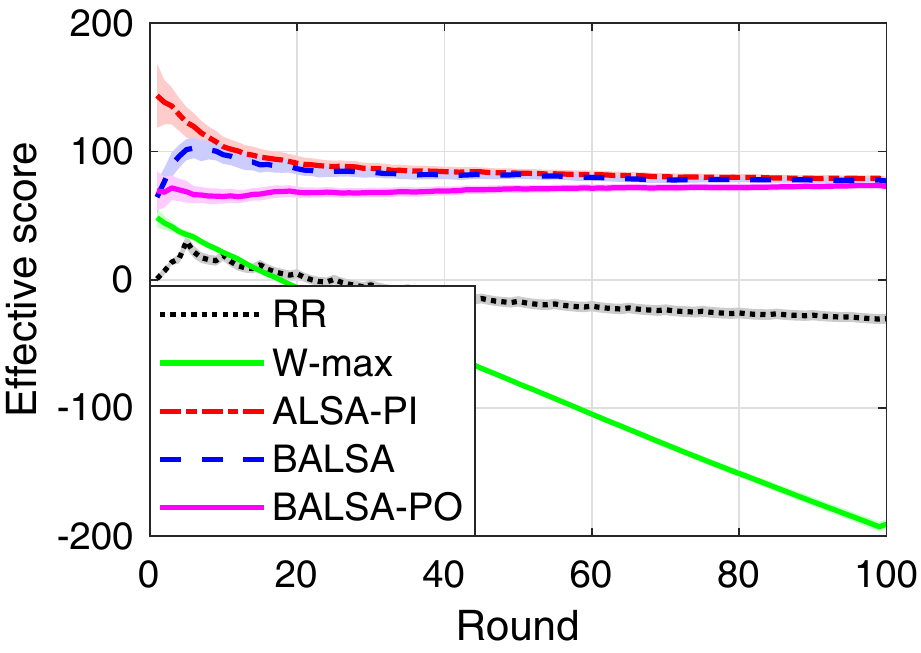}
	\caption{\label{fig:obj}Average effectivity score of the ALS problem achieved by the algorithms.}
\end{figure}

\subsection{Effectivity Scores in the ALS Problem}
We first provide the effectivity scores of the algorithms, which are the objective function of the ALS problem in \eqref{eqn:problem_ALS}, in Fig. \ref{fig:obj}.
Note that Bench is not provided in the figure since in Bench, all devices transmit their local updates and all the transmissions succeed.
From the figure, we can see that our BALSAs achieve the similar effectivity score to that of ALSA-PI, which is the optimal one.
In particular, as the round proceeds, the effectivity scores of BALSAs converge to that of ALSA-PI.
This clearly shows that our BALSAs can effectively learn the uncertainties in the WDLN.
On the other hand, RR and $W$-max achieve the much lower effectivity scores compared with BALSAs.
Especially, the effectivity score of $W$-max decreases as the round proceeds since it fails not only to maximize the number of the data samples used in training and but also to minimize the adverse effect from the stragglers.
To show the validity of the effectivity score for effective learning in asynchronous FL,
in the following subsections, it will be clearly shown that the algorithms with the higher effectivity scores achieve better trained model performances such as training loss, accuracy, robustness against stragglers, and learning speed.

\begin{figure}[!t]
	\centering
	\includegraphics[width=0.32\textwidth]{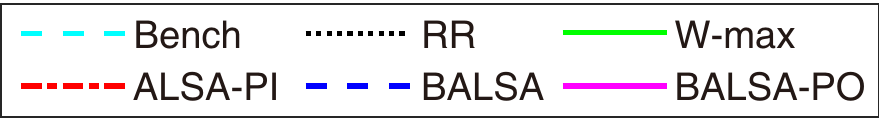} \\
	\subfloat[Training loss with MNIST]{\includegraphics[width=0.23\textwidth]{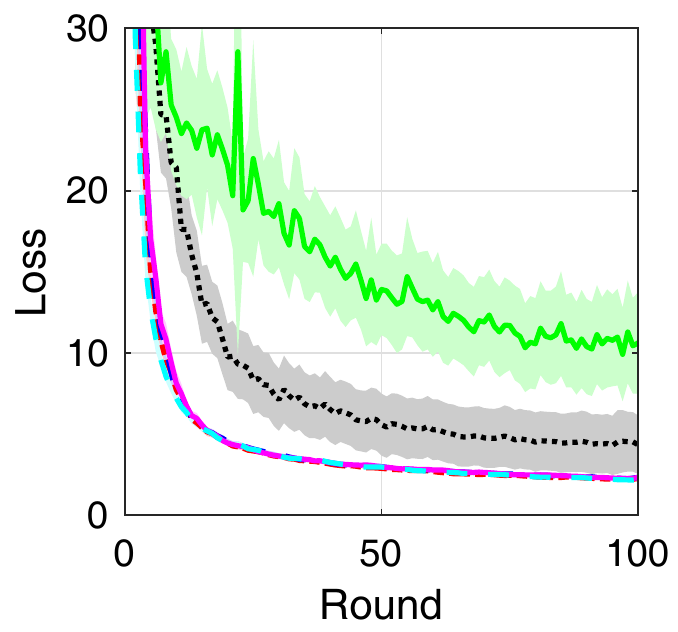}
		\label{fig:mnist_loss}}
	\hfil
	\subfloat[Test accuracy with MNIST]{\includegraphics[width=0.23\textwidth]{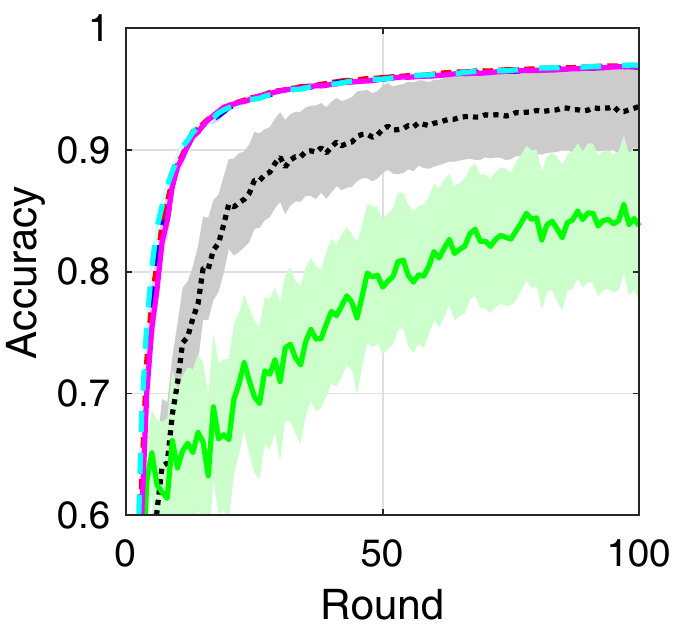}
		\label{fig:mnist_acc}}
	\hfil
	\subfloat[Training loss with CIFAR-10\looseness=-1]{\includegraphics[width=0.23\textwidth]{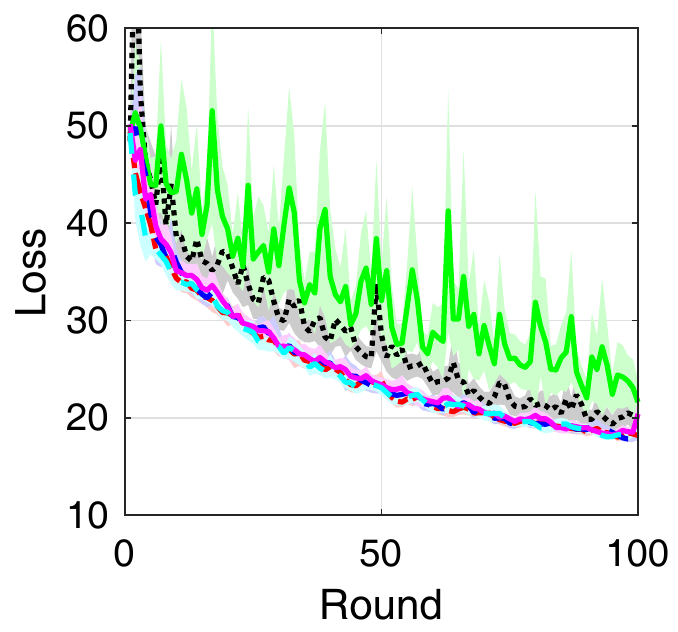}
		\label{fig:cifar10_loss}}
	\hfil
	\subfloat[Test accuracy with CIFAR-10\looseness=-1]{\includegraphics[width=0.23\textwidth]{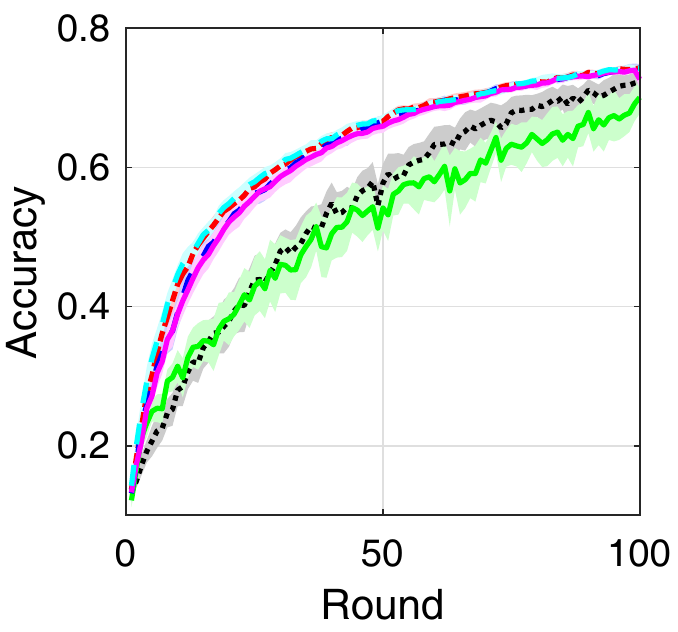}
		\label{fig:cifar10_acc}}
	\caption{Training loss and test accuracy with MNIST and CIFAR-10 datasets.}
	\label{fig:acc_loss}
\end{figure}

\subsection{Training Loss and Test Accuracy}
To compare the performance of asynchronous FL with respect to the transmission scheduling algorithms, which is our ultimate goal, we provide the training loss and test accuracy of the algorithms.
Fig. \ref{fig:acc_loss} provides the training loss and test accuracy with MNIST and CIFAR-10 datasets. From Figs. \ref{fig:mnist_loss} and \ref{fig:cifar10_loss}, we can see that ALSA-PI, BALSA, and BALSA-PO achieve the similar training loss to Bench. Compared with them, RR and $W$-max achieve the larger training loss.
The larger training loss of a model typically implies the lower accuracy of the model.
This is clearly shown in Figs. \ref{fig:mnist_acc} and \ref{fig:cifar10_acc}.
From the figures, we can see that the models trained with RR and $W$-max achieve the significantly lower accuracy than the models trained with the other algorithms and their variances are much larger than those of the other algorithms.
These results imply that RR and $W$-max fail to effectively gather the local updates while addressing the stragglers because they do not have a capability to consider the characteristics of asynchronous FL and the uncertainties in the WDLN.
On the other hand, our BALSAs gather the local updates which are enough to train the model while effectively addressing the stragglers due to asynchronous FL by considering the effectivity score.

\begin{figure}[!t]
	\centering
	\includegraphics[width=0.32\textwidth]{fig/sim/legends_nomarker_twocolumn} \\
	\subfloat[Training loss with MNIST]{\includegraphics[width=0.23\textwidth]{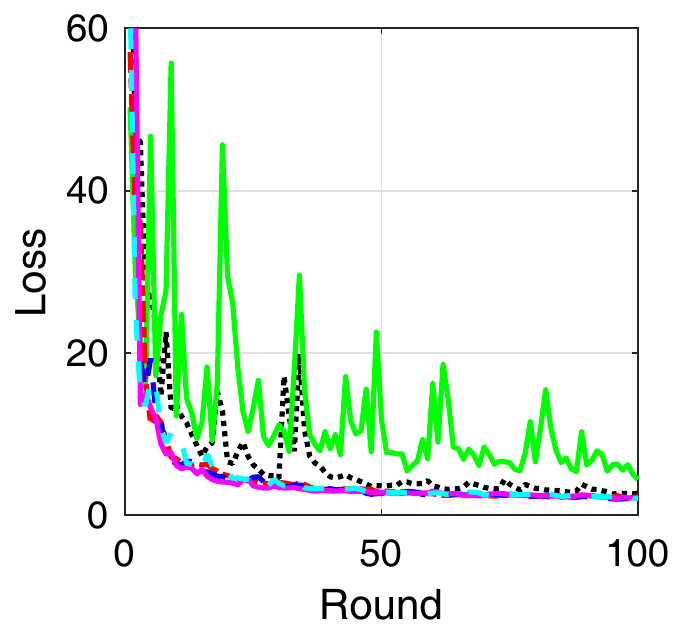}
		\label{fig:mnist_robust_loss}}
	\hfil
	\subfloat[Test accuracy with MNIST]{\includegraphics[width=0.23\textwidth]{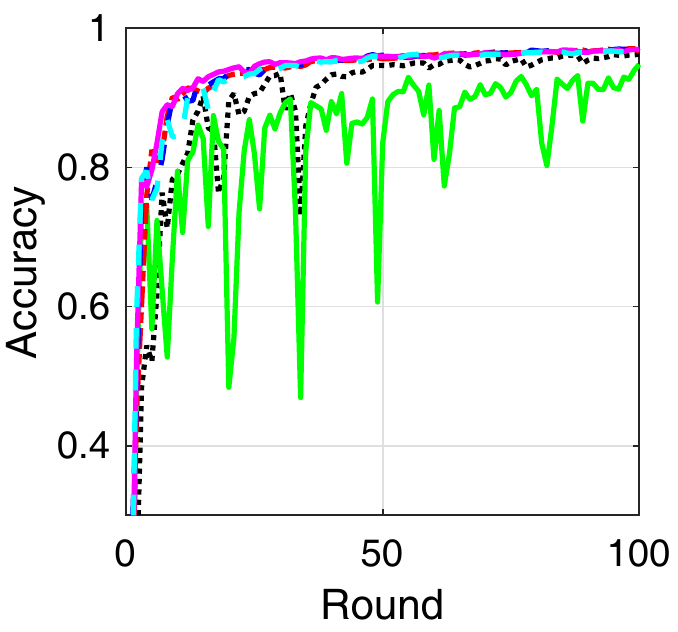}
		\label{fig:mnist_robust_acc}}
	\hfil
	\subfloat[Training loss with CIFAR-10\looseness=-1]{\includegraphics[width=0.23\textwidth]{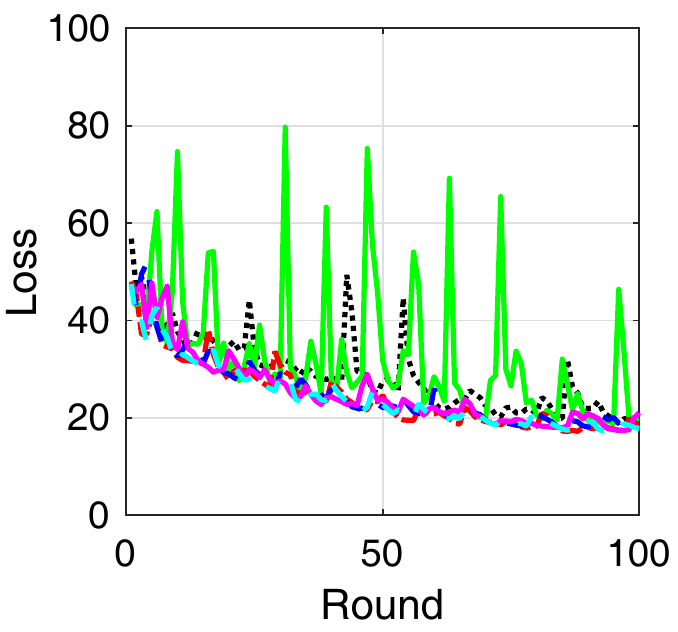}
		\label{fig:cifar10_robust_loss}}
	\hfil
	\subfloat[Test accuracy with CIFAR-10\looseness=-1]{\includegraphics[width=0.23\textwidth]{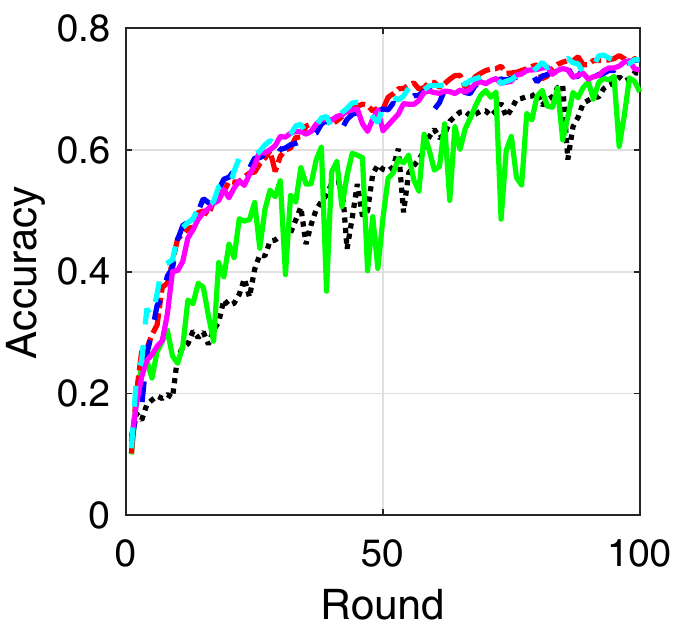}
		\label{fig:cifar10_robust_acc}}
	\caption{Training loss and test accuracy of a single simulation instance with MNIST and CIFAR-10 datasets.}
	\label{fig:robust}
\end{figure}

\subsection{Robustness of Learning against Stragglers}
In Fig. \ref{fig:acc_loss}, the unstable learning of the algorithms due to the stragglers is not clearly shown since the fluctuations of the training loss and accuracy in the figure are wiped out while averaging the results from the multiple simulation instances. Hence, in Fig. \ref{fig:robust}, we show the robustness of the algorithms against the stragglers more clearly through the training loss and test accuracy results of a single simulation instance.
First of all, it is worth emphasizing that Bench is the most stable one in terms of the stragglers because it has no straggler.
From Figs. \ref{fig:mnist_robust_loss} and \ref{fig:cifar10_robust_loss}, we can see that the training losses of the algorithms considering asynchronous FL based on the effectivity score (i.e., ALSA-PI and BALSAs) are quite stable as much as that of Bench. Accordingly, their corresponding test accuracies are also stable.
On the other hand, for the algorithms not considering asynchronous FL (i.e., RR and $W$-max), a lot of spikes (i.e., short-lasting peaks) appear in their training losses, and accordingly, their test accuracies are also unstable.
Moreover, the training loss and test accuracy of $W$-max is significantly unstable compared with RR.
This is because the scheduling strategy of $W$-max is highly biased according to the average channel gain while RR sequentially schedules all the devices. Such biased scheduling in $W$-max raises much more stragglers than RR.
This clearly shows that a transmission scheduling algorithm may cause unstable learning if its scheduling strategy is biased without considering the stragglers.\looseness=-1

\begin{figure}[!t]
	\centering
	\includegraphics[width=0.32\textwidth]{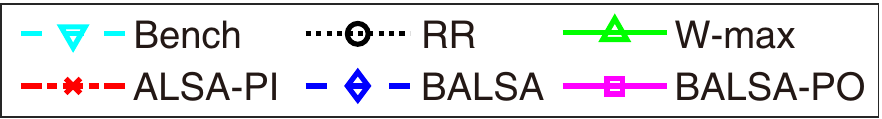} \\
	\subfloat[Target accuracy satisfaction rate with MNIST]{\includegraphics[width=0.23\textwidth]{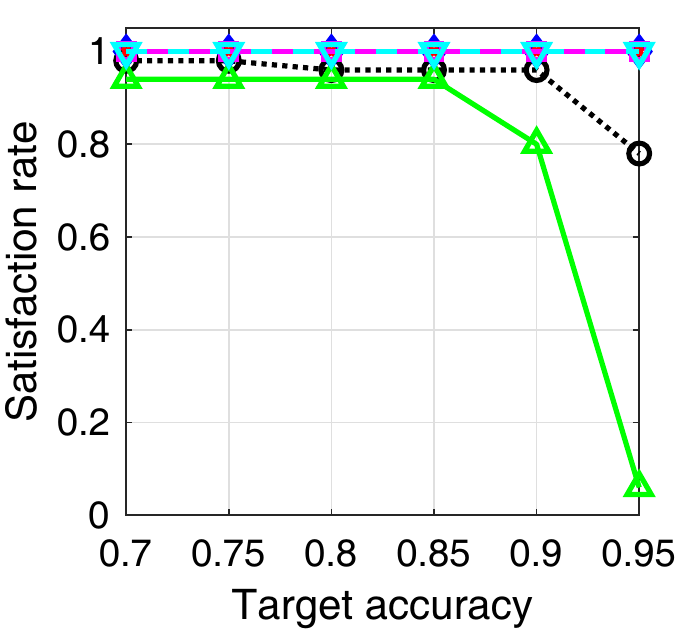}
		\label{fig:mnist_satis}}
	\hfil
	\subfloat[Average required rounds with MNIST]{\includegraphics[width=0.23\textwidth]{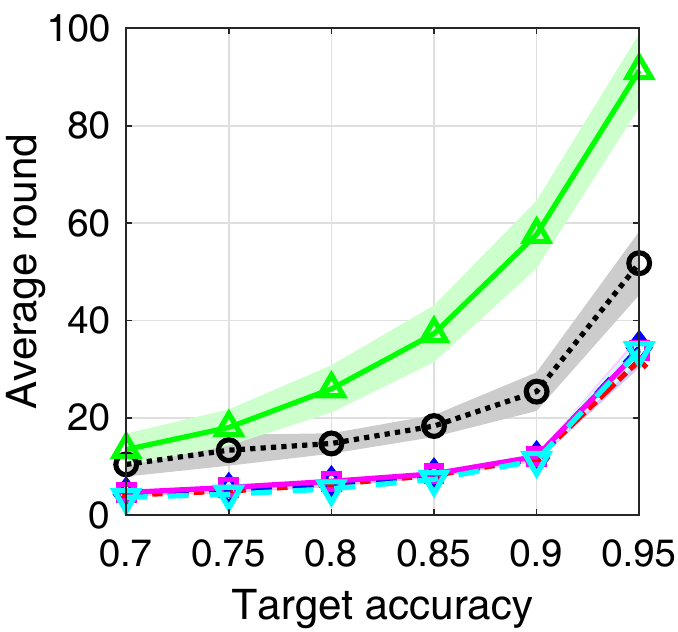}
		\label{fig:mnist_epoch}}
	\hfil
	\subfloat[Target accuracy satisfaction rate with CIFAR-10]{\includegraphics[width=0.23\textwidth]{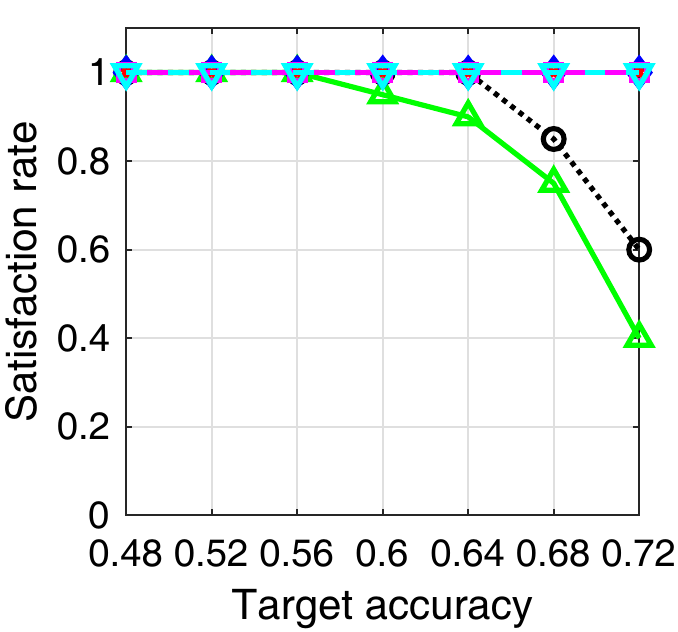}
		\label{fig:cifar10_satis}}
	\hfil
	\subfloat[Average required rounds with CIFAR-10]{\includegraphics[width=0.23\textwidth]{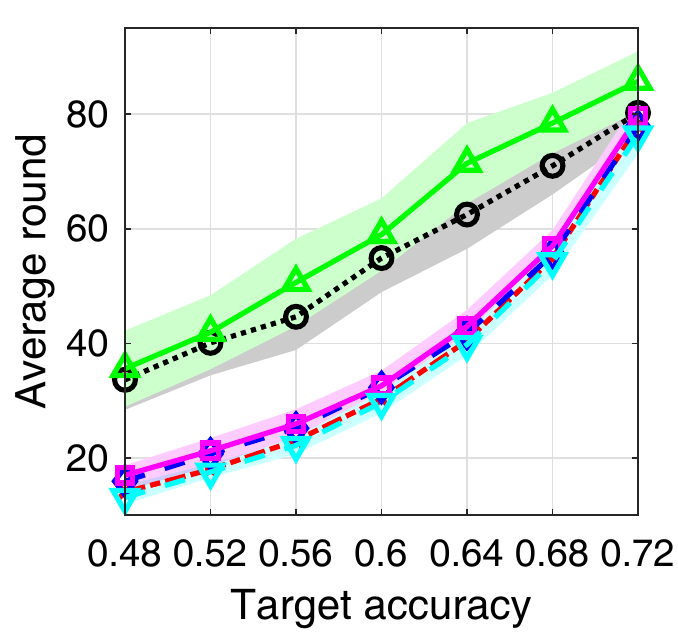}
		\label{fig:cifar10_epoch}}
	\caption{Target accuracy satisfaction rate and average required rounds for satisfying the target accuracy with MNIST and CIFAR-10 datasets.}
	\label{fig:satis_epoch}
\end{figure}

\subsection{Satisfaction Rate and Learning Speed}
In Fig. \ref{fig:satis_epoch}, we provide the target accuracy satisfaction rate and the average required rounds for satisfying the target accuracy of each algorithm.
For the MNIST dataset, we vary the target accuracy from 0.7 to 0.95, and for the CIFAR-10 dataset, we vary it from 0.48 to 0.72.
The test accuracy satisfaction rate of each algorithm is obtained as the ratio of the simulation instances, where the test accuracy of the corresponding trained model exceeds the target accuracy at the end of the simulation, to the total simulation instances.
For the average required rounds for satisfying the target accuracy, we find the minimum required rounds of each simulation instance in which the target accuracy is satisfied.
To avoid the effect of the spikes, we set a criteria of satisfying the target accuracy as follows: all test accuracies in three consecutive rounds exceed the target accuracy.
From Figs. \ref{fig:mnist_satis} and \ref{fig:cifar10_satis}, we can see that the satisfaction rates of ALSA-PI and BALSAs are similar to that of Bench.
On the other hand, the satisfaction rates of RR and $W$-max significantly decrease as the target accuracy increases. 
Figs. \ref{fig:mnist_epoch} and \ref{fig:cifar10_epoch} provide the average required rounds of the algorithms to satisfy the target accuracy.
From the figures, we can see that the algorithms considering asynchronous FL (ALSA-PI and BALSAs) require the smaller rounds to satisfy the target accuracy.
This clearly shows that the algorithms that consider the effective score of asynchronous FL (i.e., ALSA-PI and BALSAs) have a faster learning speed than the algorithms that do not consider it (i.e., RR and $W$-max).

\section{Conclusion and Future Work}
In this paper, we proposed the asynchronous FL procedure in the WDLN and investigated its convergence.
We also investigated transmission scheduling in the WDLN for effective learning.
To this end, we first proposed the effectivity score of asynchronous FL  that represents the amount of learning in which the harmful effects on learning due to the stragglers are considered.
We then formulated the ALS problem that maximizes the effectivity score of asynchronous FL.
We developed ALSA-PI that can solve the ALS problem when the perfect a priori information is given. We also developed BALSA and BALSA-PO that effectively solve the ALS problem without a priori information by learning the uncertainty on stochastic data arrivals with a minimal amount of information.
Our experimental results show that our ALSA-PI and BALSAs achieve the similar performance to the ideal benchmark. In addition, they outperform the other baseline scheduling algorithms.
These results clearly show that the transmission scheduling strategy based on the effectivity score, which is  adopted to our algorithms, is effective for asynchronous FL.
Besides, our BALSAs effectively schedules the transmissions even without any a priori information by learning the system uncertainties.
As a future work, a non-i.i.d. distribution of data samples over devices can be incorporated into the ALS problem for more effective learning in a non-i.i.d. data distribution scenario.
In addition, subchannel allocation and power control can be considered to minimize the time consumption for the FL update procedure by utilizing the resources more effectively in asynchronous FL.\looseness=-1

\appendices

\section{Proof of Theorem \ref{thm:convergence}}
\label{appendix:thm1}

We first provide the following lemma supported in \cite{chen2019asynchronous,bottou2018optimization} to prove Theorem \ref{thm:convergence}.
\begin{lemma}
\label{lemma:pf_thm1}
If $l(\bw)$ is $\xi$-strong convex, then with Assumption \ref{assumption1}, we have
\vspace{-0.1in}
\begin{equation}
2\xi(l(\bw^t)-l(\bw^*))\leq ||\nabla l(\bw^t)||^2.
\end{equation}
\end{lemma}
Using Lemma \ref{lemma:pf_thm1}, we now prove Theorem \ref{thm:convergence}.
Since $l(\bw)$ is $L$-smooth provided in Definition 1 (i.e., $f(\bx_1)-f(\bx_2)\leq \left<\nabla f(\bx_2),\bx_1-\bx_2\right>+\frac{L}{2}||\bx_1-\bx_2||^2,~\forall \bx_1,\bx_2$), the following holds:
\begin{IEEEeqnarray}{l}
l(\bw^{t+1})-l(\bw^t)\leq \left<\nabla l(\bw^t),\bw^{t+1}-\bw^t\right>+\frac{L}{2}\Vert\bw^{t+1}-\bw^t\Vert^2 \nonumber \\
=-\sum_{u\in\bar{\calU}^t}\nabla l(\bw^t)^\top \eta_u^t c_u^t \psi_u^t+\frac{L}{2}\Big\Vert\sum_{u\in\bar{\calU}^t}\eta_u^t c_u^t \psi_u^t\Big\Vert^2. \nonumber \\
\leq-\sum_{u\in\bar{\calU}^t}\nabla l(\bw^t)^\top \eta_u^t c_u^t \psi_u^t+\frac{L}{2}\Big(\sum_{u\in\bar{\calU}^t} ||\eta_u^t c_u^t \psi_u^t||\Big)^2. 
\end{IEEEeqnarray}
Let us define $q=\max_{u\in\calU}\{\eta_u^tc_u^t\}$. Then, $q >0$ and with Assumption \ref{assumption1} and gradient dissimilarity, the following holds:
\begin{IEEEeqnarray}{l}
~~\mathbb{E}[l(\bw^{t+1})]-l(\bw^t) \nonumber \\
 \leq -q \sum_{u\in\bar{\calU}^t} \nabla l(\bw^t)\mathbb{E}[\psi_u^t]+\frac{Lq^2}{2}\mathbb{E}\Big[\sum_{u\in\bar{\calU}^t}||\psi_u^t||\Big]^2 \nonumber \\
\leq -q\epsilon\sum_{u\in\calU}a_u^t x_u^t||\nabla l(\bw^t)||^2+\frac{Lq^2W^2V^2}{2}||\nabla l(\bw^t)||^2 \nonumber \\
= -q\left(\epsilon\sum_{u\in\calU}a_u^t x_u^t - \frac{LqW^2V^2}{2} \right)||\nabla l(\bw^t)||^2. \label{eqn:pf_thm1_1}
\end{IEEEeqnarray}
Let us define $\bar{\eta}^t=\max_{u\in\calU} \eta_u^t$. Then, because $c_u^t<1$ for all  $u\in\calU,t\in\mathcal{T}$, we have 
$
-q\left(\epsilon\sum_{u\in\calU}a_u^t x_u^t - \frac{LqW^2V^2}{2} \right)<-\bar{\eta}^t\left(\epsilon\sum_{u\in\calU}a_u^t x_u^t - \frac{L\bar{\eta}^tW^2V^2}{2} \right).
$
Now, with Lemma \ref{lemma:pf_thm1}, we can rewrite the inequality in \eqref{eqn:pf_thm1_1} as
\begin{IEEEeqnarray}{l}
\mathbb{E}[l(\bw^{t+1})]-l(\bw^t) \leq \\
\quad -2\xi\bar{\eta}^t \left(\epsilon\sum_{u\in\calU}a_u^t x_u^t-\frac{L\bar{\eta}^tW^2V^2}{2}\right)(l(\bw^t)-l(\bw^*)). \nonumber
\end{IEEEeqnarray}
By subtracting $l(\bw^*)$ from both sides and rearranging the inequality, we have\looseness=-1
\begin{IEEEeqnarray}{l}
\mathbb{E}[l(\bw^{t+1})]-l(\bw^*) \leq  \\
\quad\left\lbrace 1 +2\xi\bar{\eta}^t \left(\frac{L\bar{\eta}^tW^2V^2}{2} - \epsilon\sum_{u\in\calU}a_u^t x_u^t\right)\right\rbrace(l(\bw^t)-l(\bw^*)). \nonumber
\end{IEEEeqnarray}
By taking expectation of both sides, we can get
\begin{IEEEeqnarray}{l}
\mathbb{E}[l(\bw^{t+1})-l(\bw^*)] \leq  \label{eqn:thm1_proof_1} \\
\left\lbrace 1 +2\xi\bar{\eta}^t \left(\frac{L\bar{\eta}^tW^2V^2}{2} - \epsilon\!\sum_{u\in\calU}a_u^t \mathbb{P}[x_u^t\!=\!1]\right)\right\rbrace(l(\bw^t)-l(\bw^*)),\nonumber
\end{IEEEeqnarray}
and with telescoping the above equations, we have the equation in \eqref{eqn:thm1_eq1} in Theorem \ref{thm:convergence}.
Then, if $\sum_{u\in\calU}a_u^t \mathbb{P}[x_u^t=1] \geq \zeta$, we can rewrite the equation in \eqref{eqn:thm1_proof_1} as
\begin{equation}
\mathbb{E}[l(\bw^{t+1})-l(\bw^*)] \leq (1 -2\xi\underline{\eta} \epsilon')(l(\bw^t)-l(\bw^*)),
\end{equation}
where $\epsilon'=\epsilon\zeta-\frac{L\bar{\eta} W^2V^2}{2}$ and $\bar{\eta}=\frac{2\epsilon\zeta}{LW^2V^2}(\max_{u\in\calU,\forall t}\{c_u^t\})^{-1}$.
With telescoping the above equations, we have the equation in \eqref{eqn:thm1_eq2} in Theorem \ref{thm:convergence}.

\section{Proof of Theorem \ref{thm:greedy_optimal}}
\label{appendix:thm2}
Suppose that there exists the optimal policy whose chosen action in a round with state $\bs$ is not identical to the greedy policy.
We denote the expected instantaneous effectivity score of round $t$ with the optimal policy by $r_*^t(\bs)$ and that with the greedy policy by $r_g^t(\bs)$.
Based on the equation in \eqref{eqn:objective_function2}, we can decompose the expected instantaneous effectivity score into two sub-rewards as $r(\bs)=r_A(\bs)+r_B(\bs)$, where $r_A(\bs)=\mathbb{E}[\sum_{u\in\calU} a_u x_u(1+\gamma)(n_u+m_u)]$ and $r_B(\bs)=-\mathbb{E}[\sum_{u\in\calU} \gamma(n_u+m_u)]$.
In the problem, the samples are accumulated if they are not used for central training.
Moreover, the channel gains and arrival rates of the samples of the devices are i.i.d.
Hence, if a policy $\pi$ satisfies\looseness=-1
\begin{equation}
\label{eqn:thm2:cond1}
\sum_{s\in\calS}\mathbb{P}[a_u^\pi(\bs)]>0 \textrm{ for all }u\in\calU,
\end{equation} where $a_u^\pi$ represents $a_u$ from policy $\pi$,
any arrived samples will be eventually reflected in the sub-reward $r_A$ regardless of the policy due to the accumulation of the samples.
This implies that the average of sub-reward $\sum_{t=1}^T r_A^t/T$ converges to $\sum_{u\in\calU}\theta_u^P$ as $T\rightarrow\infty$.
Accordingly, if a policy satisfies the condition in \eqref{eqn:thm2:cond1}, its optimality to the ALS problem depends on minimizing the delay cost (i.e., the sub-reward $r_B$).
With the greedy policy, any device will be scheduled when the number of its aggregated samples is large enough.
Hence, the greedy policy satisfies the condition if the parameters are finite.
We now denote the expected sub-rewards $r_B$ of round $t$ with the optimal policy and the greedy policy by $r_{B,*}^{t}$ and $r_{B,g}^{t}$, respectively.
From the definition of the greedy policy, it is obvious that
$
\mathbb{E}[\sum_{u\in\calU}n_u^{t}|\ba_*]\geq \mathbb{E}[\sum_{u\in\calU}n_u^{t}|\ba_g],
$
which implies that
$
r_{B,*}^{t}\leq r_{B,g}^{t},
$
for any random disturbances.
Consequently, this leads to
$
J^*\leq J^g,
$
which implies that the greedy policy is optimal.

\section{Proof of Theorem \ref{thm:regret_bound}}
\label{appendix:thm3}
We define $K_T=\argmax\{k:t_k\leq T\}$, which represents the number of stages in BALSA until round $T$.
For $t_k\leq t< t_{k+1}$ in stage $k$, we have the following equation from  \eqref{eqn:bellman_equation}:
\begin{equation}
r(\bs^t,\ba^t)=J(\btheta_k)+v(\bs^t,\btheta_k)-\sum_{\bs'\in\calS}P^{\btheta_k}(\bs'|\bs^t,\ba^t)v(\bs',\btheta_k).
\end{equation}
Then, the expected regret of BALSA is derived as
\begin{IEEEeqnarray}{l}
\label{eqn:regret_theorem}
T\mathbb{E}[J(\btheta_{*})]-\mathbb{E}\Big[\sum_{k=1}^{K_T}\sum_{t=t_k}^{t_{k+1}-1}r(\bs^t,\ba^t)\Big] = R_1+R_2+R_3,
\end{IEEEeqnarray}
where
\begin{IEEEeqnarray}{l}
R_1=T\mathbb{E}[J(\btheta_{*})] \!-\! \mathbb{E}\Big[\sum_{k=1}^{K_T}T_k J(\btheta_k)\Big], \nonumber \\
R_2=\mathbb{E}\Big[\sum_{k=1}^{K_T}\sum_{t=t_k}^{t_{k+1}-1} v(\bs^{t+1},\btheta_k)- v(\bs^t,\btheta_k)\Big], \nonumber \\
\textrm{and }R_3=\mathbb{E}\Big[\sum_{k=1}^{K_T}\sum_{t=t_k}^{t_{k+1}-1} \!\sum_{\bs'\in\calS}P^{\btheta_k}(\bs'|\bs^t,\ba^t)v(\bs',\btheta_k)\!-\!v(\bs^{t+1},\btheta_k)\Big].\nonumber
\end{IEEEeqnarray}
We can bound the regret of BALSA by deriving the bounds on $R_1$, $R_2$, and $R_3$ as the following lemma:
\begin{lemma}
\label{lemma:regret}
For the expected regret of BALSA, we have the following bounds:
\begin{itemize}
\item The first term is bounded as $R_1\leq \mathbb{E}[K_T]$.
\item The second term is bounded as $R_2\leq \mathbb{E}[HK_T]$.
\item The third term is bounded as $R_3\leq 49HS\sqrt{AT\log(AT)}$.
\end{itemize}
\end{lemma}
In addition, we can bound the number of stages $K_T$ as follows.
\begin{lemma}
\label{lemma:stage_number}
The number of stages in BALSA until round $T$ is bounded as $K_T\leq \sqrt{2SAT\log(T)}$.
\end{lemma}
The above lemmas can be proven in similar steps to Lemmas 1--5 in \cite{ouyang2017learning}.
Hence, here we omit the proofs due to the lack of space. For more details, we refer to \cite{ouyang2017learning}.
From the equation in \eqref{eqn:regret_theorem}, we have $R(T)=R_1+R_2+R_3$.
Then, Theorem \ref{thm:regret_bound} holds by Lemmas \ref{lemma:regret} and \ref{lemma:stage_number}.

\bibliographystyle{IEEEtran}
\bibliography{IEEEabrv,mybib}

\begin{thebibliography}{10}
\providecommand{\url}[1]{#1}
\csname url@samestyle\endcsname
\providecommand{\newblock}{\relax}
\providecommand{\bibinfo}[2]{#2}
\providecommand{\BIBentrySTDinterwordspacing}{\spaceskip=0pt\relax}
\providecommand{\BIBentryALTinterwordstretchfactor}{4}
\providecommand{\BIBentryALTinterwordspacing}{\spaceskip=\fontdimen2\font plus
\BIBentryALTinterwordstretchfactor\fontdimen3\font minus
  \fontdimen4\font\relax}
\providecommand{\BIBforeignlanguage}[2]{{%
\expandafter\ifx\csname l@#1\endcsname\relax
\typeout{** WARNING: IEEEtran.bst: No hyphenation pattern has been}%
\typeout{** loaded for the language `#1'. Using the pattern for}%
\typeout{** the default language instead.}%
\else
\language=\csname l@#1\endcsname
\fi
#2}}
\providecommand{\BIBdecl}{\relax}
\BIBdecl

\bibitem{mcmahan2017communication}
B.~McMahan, E.~Moore, D.~Ramage, S.~Hampson, and B.~A. y~Arcas,
  ``Communication-efficient learning of deep networks from decentralized
  data,'' in \emph{Proc. AISTATS}, 2017.

\bibitem{lim2020federated}
W.~Y.~B. Lim, N.~C. Luong, D.~T. Hoang, Y.~Jiao, Y.-C. Liang, Q.~Yang,
  D.~Niyato, and C.~Miao, ``Federated learning in mobile edge networks: A
  comprehensive survey,'' \emph{{IEEE} Commun. Surveys Tuts.}, no.~3, pp.
  2031--2063, 2020.

\bibitem{amiri2020federated}
M.~M. Amiri and D.~G{\"u}nd{\"u}z, ``Federated learning over wireless fading
  channels,'' \emph{{IEEE} Trans. Wireless Commun.}, vol.~19, no.~5, pp.
  3546--3557, May 2020.

\bibitem{amiri2020machine}
------, ``Machine learning at the wireless edge: Distributed stochastic
  gradient descent over-the-air,'' \emph{{IEEE} Trans. Signal Process.},
  vol.~68, pp. 2155--2169, Mar. 2020.

\bibitem{yang2019scheduling}
H.~H. Yang, Z.~Liu, T.~Q. Quek, and H.~V. Poor, ``Scheduling policies for
  federated learning in wireless networks,'' \emph{{IEEE} Trans. Commun.},
  vol.~68, no.~1, pp. 317--333, Jan. 2020.

\bibitem{wang2019adaptive}
S.~Wang, T.~Tuor, T.~Salonidis, K.~K. Leung, C.~Makaya, T.~He, and K.~Chan,
  ``Adaptive federated learning in resource constrained edge computing
  systems,'' \emph{{IEEE} J. Sel. Areas Commun.}, vol.~37, no.~6, pp.
  1205--1221, June 2019.

\bibitem{amiriconvergence}
M.~M. Amiri, D.~G{\"u}nd{\"u}z, S.~R. Kulkarni, and H.~V. Poor, ``Convergence
  of update aware device scheduling for federated learning at the wireless
  edge,'' vol.~20, no.~6, pp. 3643--3658, June 2021.

\bibitem{shi2020joint}
W.~Shi, S.~Zhou, Z.~Niu, M.~Jiang, and L.~Geng, ``Joint device scheduling and
  resource allocation for latency constrained wireless federated learning,''
  \emph{{IEEE} Trans. Wireless Commun.}, vol.~20, no.~1, pp. 453--467, Jan.
  2021.

\bibitem{chen2020joint}
M.~Chen, Z.~Yang, W.~Saad, C.~Yin, H.~V. Poor, and S.~Cui, ``A joint learning
  and communications framework for federated learning over wireless networks,''
  \emph{{IEEE} Trans. Wireless Commun.}, vol.~20, no.~1, pp. 269--283, Jan.
  2021.

\bibitem{xu2020client}
J.~Xu and H.~Wang, ``Client selection and bandwidth allocation in wireless
  federated learning networks: A long-term perspective,'' \emph{{IEEE} Trans.
  Wireless Commun.}, vol.~20, no.~2, pp. 1188--1200, Feb. 2021.

\bibitem{xia2020multi}
W.~Xia, T.~Q. Quek, K.~Guo, W.~Wen, H.~H. Yang, and H.~Zhu, ``Multi-armed
  bandit-based client scheduling for federated learning,'' vol.~19, no.~11, pp.
  7108--7123, Nov. 2020.

\bibitem{chen2019communication}
Y.~Chen, X.~Sun, and Y.~Jin, ``Communication-efficient federated deep learning
  with layerwise asynchronous model update and temporally weighted
  aggregation,'' \emph{{IEEE} Trans. Neural Netw. Learn. Syst.}, vol.~31,
  no.~10, pp. 4229--4238, Oct. 2020.

\bibitem{chen2019asynchronous}
Y.~Chen, Y.~Ning, M.~Slawski, and H.~Rangwala, ``Asynchronous online federated
  learning for edge devices with non-{IID} data,'' in \emph{Proc. 2020 IEEE
  Int. Conf. on Big Data (Big Data)}, 2020.

\bibitem{zheng17b}
S.~Zheng, Q.~Meng, T.~Wang, W.~Chen, N.~Yu, Z.-M. Ma, and T.-Y. Liu,
  ``Asynchronous stochastic gradient descent with delay compensation,'' in
  \emph{Proc. ICML}, 2017.

\bibitem{xie2019asynchronous}
C.~Xie, S.~Koyejo, and I.~Gupta, ``Asynchronous federated optimization,''
  \emph{arXiv preprint arXiv:1903.03934}, 2020.

\bibitem{chen2018lag}
T.~Chen, G.~Giannakis, T.~Sun, and W.~Yin, ``{LAG}: Lazily aggregated gradient
  for communication-efficient distributed learning,'' in \emph{Proc. NIPS},
  2018.

\bibitem{lin2018deep}
Y.~Lin, S.~Han, H.~Mao, Y.~Wang, and B.~Dally, ``Deep gradient compression:
  Reducing the communication bandwidth for distributed training,'' in
  \emph{Proc. ICLR}, 2018.

\bibitem{xu2020ternary}
J.~Xu, W.~Du, Y.~Jin, W.~He, and R.~Cheng, ``Ternary compression for
  communication-efficient federated learning,'' \emph{{IEEE} Trans. Neural
  Netw. Learn. Syst.}, to be published.

\bibitem{li2020multi}
F.~Li, D.~Yu, H.~Yang, J.~Yu, H.~Karl, and X.~Cheng, ``Multi-armed-bandit-based
  spectrum scheduling algorithms in wireless networks: A survey,'' \emph{{IEEE}
  Wireless Commun.}, vol.~27, no.~1, pp. 24--30, 2020.

\bibitem{lee2019resource}
H.-S. Lee, J.-Y. Kim, and J.-W. Lee, ``Resource allocation in wireless networks
  with deep reinforcement learning: A circumstance-independent approach,''
  \emph{{IEEE} Syst. J.}, vol.~14, no.~2, pp. 2589--2592, 2020.

\bibitem{smith2017federated}
V.~Smith, C.-K. Chiang, M.~Sanjabi, and A.~S. Talwalkar, ``Federated multi-task
  learning,'' in \emph{Proc. NIPS}, 2017.

\bibitem{baytas2016asynchronous}
I.~M. Baytas, M.~Yan, A.~K. Jain, and J.~Zhou, ``Asynchronous multi-task
  learning,'' in \emph{IEEE ICDM}, 2016.

\bibitem{xi2011general}
Y.~Xi, A.~Burr, J.~Wei, and D.~Grace, ``A general upper bound to evaluate
  packet error rate over quasi-static fading channels,'' \emph{{IEEE} Trans.
  Wireless Commun.}, vol.~10, no.~5, pp. 1373--1377, May 2011.

\bibitem{ferrand2015approximations}
P.~Ferrand, J.-M. Gorce, and C.~Goursaud, ``Approximations of the packet error
  rate under quasi-static fading in direct and relayed links,'' \emph{EURASIP
  J. on Wireless Commun. and Netw.}, vol. 2015, no.~1, p.~12, Jan. 2015.

\bibitem{wu2013energy}
J.~Wu, G.~Wang, and Y.~R. Zheng, ``Energy efficiency and spectral efficiency
  tradeoff in type-{I} {ARQ} systems,'' \emph{{IEEE} J. Sel. Areas Commun.},
  vol.~32, no.~2, pp. 356--366, Feb. 2014.

\bibitem{ge2014packet}
S.~Ge, Y.~Xi, S.~Huang, and J.~Wei, ``Packet error rate analysis and power
  allocation for cc-harq over rayleigh fading channels,'' \emph{{IEEE} Commun.
  Lett.}, vol.~18, no.~8, pp. 1467--1470, Aug. 2014.

\bibitem{MLSYS2020_38af8613}
T.~Li, A.~K. Sahu, M.~Zaheer, M.~Sanjabi, A.~Talwalkar, and V.~Smith,
  ``Federated optimization in heterogeneous networks,'' in \emph{Proceedings of
  Machine Learning and Systems}, vol.~2, 2020, pp. 429--450.

\bibitem{216799}
Z.~Tao and Q.~Li, ``{eSGD}: Communication efficient distributed deep learning
  on the edge,'' in \emph{Proc. {USENIX} Workshop Hot Topics Edge Comput.
  (HotEdge 18)}, 2018.

\bibitem{domingos2012few}
P.~Domingos, ``A few useful things to know about machine learning,''
  \emph{Commun. of the ACM}, vol.~55, no.~10, pp. 78--87, 2012.

\bibitem{goodfellow2016deep}
I.~Goodfellow, Y.~Bengio, A.~Courville, and Y.~Bengio, \emph{Deep
  learning}.\hskip 1em plus 0.5em minus 0.4em\relax MIT press Cambridge, 2016,
  vol.~1, no.~2.

\bibitem{bertsekas1995dynamic}
D.~P. Bertsekas, \emph{Dynamic programming and optimal control}.\hskip 1em plus
  0.5em minus 0.4em\relax Athena scientific Belmont, MA, 1995, vol.~1, no.~2.

\bibitem{gopalan2015thompson}
A.~Gopalan and S.~Mannor, ``Thompson sampling for learning parameterized
  {Markov} decision processes,'' in \emph{Proc. Conf. on Learn. Theory}, 2015.

\bibitem{osband2017posterior}
I.~Osband and B.~Van~Roy, ``Why is posterior sampling better than optimism for
  reinforcement learning?'' in \emph{Proc. ICML}, 2017.

\bibitem{ouyang2017learning}
Y.~Ouyang, M.~Gagrani, A.~Nayyar, and R.~Jain, ``Learning unknown {Markov}
  decision processes: A {Thompson} sampling approach,'' in \emph{Proc. NIPS},
  2017.

\bibitem{lee2021sdf}
H.-S. Lee, C.~Shen, W.~Zame, J.-W. Lee, and M.~Schaar, ``{SDF-Bayes}: Cautious
  optimism in safe dose-finding clinical trials with drug combinations and
  heterogeneous patient groups,'' in \emph{Proc. AISTATS}, 2021.

\bibitem{misgiyati2017bayesian}
M.~Misgiyati and K.~Nisa, ``Bayesian inference of poisson distribution using
  conjugate and non-informative priors,'' in \emph{Prosiding Seminar Nasional
  Metode Kuantitatif}, no.~1, 2017.

\bibitem{hou2011demodulation}
X.~Hou and H.~Kayama, \emph{Demodulation reference signal design and channel
  estimation for LTE-Advanced uplink}.\hskip 1em plus 0.5em minus 0.4em\relax
  INTECH Open Access Publisher, 2011.

\bibitem{Simonyan15}
K.~Simonyan and A.~Zisserman, ``Very deep convolutional networks for
  large-scale image recognition,'' in \emph{Proc. ICLR}, 2015.

\bibitem{bottou2018optimization}
L.~Bottou, F.~E. Curtis, and J.~Nocedal, ``Optimization methods for large-scale
  machine learning,'' \emph{Siam Review}, vol.~60, no.~2, pp. 223--311, 2018.

\end{thebibliography}
	
\end{document}